# On Constant-Round Concurrent Zero-Knowledge from a Knowledge Assumption

Divya Gupta    Amit Sahai
{divyag, sahai}@cs.ucla.edu
University of California, Los Angeles, USA

**Abstract**

In this work, we consider the long-standing open question of constructing constant-round concurrent zero-knowledge protocols in the plain model. Resolving this question is known to require non-black-box techniques.

We consider non-black-box techniques for zero-knowledge based on knowledge assumptions, a line of thinking initiated by the work of Hada and Tanaka (CRYPTO 1998). Prior to our work, it was not known whether knowledge assumptions could be used for achieving security in the concurrent setting, due to a number of significant limitations that we discuss here. Nevertheless, we obtain the following results:

1. We obtain the first constant round concurrent zero-knowledge argument for **NP** in the plain model based on a new variant of knowledge of exponent assumption. Furthermore, our construction avoids the inefficiency inherent in previous non-black-box techniques such that those of Barak (FOCS 2001); we obtain our result through an efficient protocol compiler.

2. Unlike Hada and Tanaka, we do not require a knowledge assumption to argue the soundness of our protocol. Instead, we use a discrete log like assumption, which we call *Diffie-Hellman Logarithm Assumption*, to prove the soundness of our protocol.

3. We give evidence that our new variant of knowledge of exponent assumption is in fact plausible. In particular, we show that our assumption holds in the generic group model.

4. Knowledge assumptions are especially delicate assumptions whose plausibility may be hard to gauge. We give a novel framework to express knowledge assumptions in a more flexible way, which may allow for formulation of plausible assumptions and exploration of their impact and application in cryptography.

**Keywords.** Concurrent Zero-Knowledge, Knowledge Assumptions, Non-Black-Box Techniques

# 1 Introduction

Zero-knowledge proofs [GMR89] are fundamental and important tools in the design of cryptographic protocols. The original setting of zero-knowledge proofs contemplated a single prover and a single verifier executing a single protocol session in isolation. Concurrent zero-knowledge [DNS98] ($c\mathcal{ZK}$) extends zero-knowledge to concurrent settings, where several protocol sessions are executed at the same time involving multiple provers and verifiers. Resolving the round complexity of concurrent zero-knowledge protocols has been a long standing open problem. There have been several negative results which give lower bounds for round complexity of black box simulation of $c\mathcal{ZK}$ [KPR98, Ros00, CKPR01]. The best result, which uses black box simulation, has $\omega(\log n)$ round complexity [PRS02], where $n$ is the security parameter. This nearly matches the best known lower bound for black box simulation [CKPR01], which states that any black-box concurrent zero-knowledge protocol must require at least $\tilde{\Omega}(\log n)$ rounds. Hence, our only hope of achieving constant round $c\mathcal{ZK}$ lies in non-black-box simulation. In his seminal work, Barak [Bar01] introduced the first non-black-box simulation technique, but this technique or its variants have not yet yielded a concurrent zero-knowledge protocol with lower round complexity than the work of [PRS02]. Indeed, Barak explicitly posed the problem of constructing constant-round concurrent zero-knowledge arguments as "an important open question" [Bar01]. Despite many attempts in this direction, this is still a long-standing open problem in cryptography.

In this work, we consider whether non-black-box techniques based on knowledge assumptions can be applied to achieve constant round concurrent zero-knowledge protocols. We answer this question affirmatively, by giving the first constant-round concurrent zero-knowledge protocol based on a knowledge assumption, which is a novel variant of the knowledge of exponent assumption first introduced by Damgard [Dam91] and used by Hada and Tanaka [HT98] in the context of ordinary zero-knowledge protocols.

Furthermore, our techniques allow us to avoid the inefficiency inherent in previous non-black-box techniques, such as those of Barak [Bar01]. Indeed, we obtain our result by providing an efficient transformation from constant round stand alone protocols to constant round concurrently secure zero-knowledge protocols.

Recently, there has been an explosion of interest in knowledge assumptions. Knowledge assumptions became popular when these were applied to the construction of three round zero-knowledge arguments by [HT98]. This has led to a number of interesting research papers applying knowledge assumptions to a variety of different contexts [BP04, AF07, CD08, CL08, CD09, PX09, IKOS10, Gro10, GKR10, GLR11, BCCT12, DFH12]. Prior to our work, to the best of our knowledge, knowledge assumptions have not been applied successfully to achieve concurrent security. As we explore below, this is because of a number of complications which arise when one applies knowledge assumptions to concurrent settings.

**Our Contributions:** We show the following:

1. We obtain the first constant round concurrent zero-knowledge argument for **NP** in plain model based on a new variant of knowledge of exponent assumption. Our compiler to get concurrently secure protocol is efficient and avoids the inefficiency inherent in other non-black-box techniques.

2. Unlike Hada and Tanaka [HT98], we do not require a knowledge assumption to argue the soundness of our protocol. Instead, we use a discrete log like assumption, which we call *DHLA* (See Section 3.1), to prove the soundness of our protocol.

3. We give evidence that our new variant of knowledge of exponent assumption is in fact plausible. In particular, we show that our assumption holds in the generic group model.

4. As we discuss in greater detail below, and as has been discussed throughout the history of knowledge assumptions in cryptography, knowledge assumptions are especially delicate assumptions whose plausibility may be hard to gauge. We give a novel framework to express knowledge assumptions in a more flexible way which may allow for formulation of plausible assumptions and exploration of their impact and application in cryptography.

**On Knowledge Assumptions and their Applications in Cryptography.** Knowledge assumptions are inherently non-black-box. Informally speaking[1], knowledge assumptions can be expressed by assuming that there is a specific "Knowledge Commitment Protocol" such that we can efficiently extract the value committed by the adversary if he completes the commitment protocol successfully — in other words, we assume that any adversary that successfully completes the Knowledge Commitment Protocol must have "knowledge" of the value that it committed to. For the purpose of this introduction, assume that the Knowledge Commitment Protocol is just a two message protocol in which first the Receiver sends a random message to the Committer and then the Committer responds with a commitment to a value[2]. Knowledge assumptions present a number of challenges to the research community from the point of view of the falsifiability rubric of Naor [Nao03]: they do not fall in the desirable category of falsifiable assumptions in [Nao03].

Furthermore, knowledge assumptions present challenges with regard to auxiliary inputs as is also pointed out in the early works of Hada and Tanaka [HT98]. Intuitively the problem arises if we consider what happens if an adversary is given as auxiliary input an obfuscated program. The adversary simply compiles and executes the obfuscated program to obtain the commitment message. Then a knowledge assumption, which is expected to hold *for all* auxiliary inputs, would imply an efficient extraction of the committed value. This would imply an efficient deobfuscation, which seems problematic. It was recently suggested by Bitansky et al [BCCT12] that it is more reasonable to assume that knowledge assumptions only hold with respect to "benign" auxiliary inputs. One of our contributions is to put forward a framework for formulating knowledge assumptions with respect to *Admissible Adversaries*. This allows us to specify a set of auxiliary inputs with respect to which the knowledge of exponent assumption would hold. For applications in cryptography we want this class to be as large as possible. Despite these drawbacks, the study of knowledge assumptions in cryptography has been thriving recently. This is evident by the long list of interesting research papers cited above. (See Section 8 for more details).

**Limitations of Knowledge Assumptions in the Setting of Concurrency.** Undoubtedly, the reason that knowledge assumptions have attracted attention is because they are very useful to achieve important goals in cryptography. Indeed often it may seem that knowledge assumptions are so powerful that they can be used to achieve any plausible result that we want to achieve in cryptography. For example, when it comes to the simulation of protocols, intuitively it seems that whenever the adversary commits to some value, the simulator can use the knowledge assumption to extract the hidden value committed to. Hence, it seems this can become a universal technique for straight line simulation[3]. This intuition is false, as we describe below.

One way to see that the above intuition is false is by observing a long list of unconditional impossibility results for concurrent simulations in plain model [CF01, Lin03, Lin04, BPS06, KLR10, GKOV12, AGJ+12] and observing that the above intuition seems to give a simulation technique applicable to any concurrent setting. Even in restricted models of concurrency, there are many natural protocol tasks that are impossible even with knowledge assumptions. One of the most relevant examples is concurrently secure oblivious transfer (OT), in the "fixed roles" setting and with fixed inputs for honest parties. This setting is almost identical to concurrent zero-knowledge, with the only difference being that there one is trying to achieve OT as opposed to zero-knowledge, but there are no issues of "man-in-the-middle attacks" or adaptive choice of inputs. Nevertheless, a concurrently secure OT protocol in fixed roles setting and with fixed inputs for honest parties is impossible even with knowledge assumptions [GKOV12,

---

[1]Our assumption is concrete. See Section 3.4.

[2]The commitment here does not refer to a semantically secure commitment scheme.

[3]For example, consider the following coin flipping protocol. Adversary commits to $R$, honest party sends $R'$, adversary opens $R$. The result of coin flipping protocol would be $R \oplus R'$. Intuitively, knowledge assumption would allow the simulator to force the outcome of coin flipping to any string he wants since it would know R immediately after the adversary's commitment through extraction. Thus, one might conclude that with knowledge assumptions we can achieve the CRS model. This intuition is false.

AGJ+12], yet as we show, there is a plausible assumption under which we achieve constant round concurrent zero-knowledge. The negative results show that potential specific knowledge assumptions, which would be powerful enough to allow for concurrently secure OT, *must be false*. (We stress that the novel knowledge of exponent assumption we formulate here would *not* naturally provide a simulation of a concurrent OT protocol.)

As suggested above, one of the main non-trivialities of our work is to formulate a plausible knowledge assumption that would allow us to achieve constant round concurrent zero-knowledge, while remaining plausible. We begin our discussion here with natural attempts to apply knowledge assumptions to the concurrent setting, and their limitations. We believe that this discussion will be useful to other researchers who would like to apply knowledge assumptions to other interesting problems in cryptography, while also illustrating the non-triviality of achieving concurrent security using a plausible knowledge assumption.

Perhaps the most promising idea would be to formulate an "interactive" knowledge assumption. Informally speaking, such an assumption would say that extraction is possible after an arbitrary interaction which took place prior to the final message in the Knowledge Commitment Protocol. However, any natural formulation of such an interactive knowledge assumption would be powerful enough to achieve concurrent realization of functionalities such as OT. Hence, we know that such an assumption must be false. Indeed such an assumption would be falsified by considering a scenario in which the actions of the adversary in the Knowledge Commitment Protocol are fully specified by messages that the adversary received in the past, and not directly by the adversary itself. (For example, the functionality being computed could provide the messages of the Knowledge Commitment Protocol as outputs to the adversary [BPS06, AGJ+12].) Intuitively, in such a situation, the adversary doesn't have any knowledge of the value he committed to, and hence the goal of extraction is untenable. Essentially the problem is that some "external knowledge" may find its way to the adversary by means of previous interactions and get used by it to generate its messages in Knowledge Commitment Protocol. Similar problems arise when trying to use auxiliary inputs to the extractor promised by a knowledge assumption in order to facilitate extraction in the concurrent setting. (See Appendix A for a brief discussion.)

**Recursive Applications of Knowledge Assumptions and their Limitations.**
Another approach would be to apply a knowledge assumption *recursively* for each session. What we mean by this is as follows: Essentially, a knowledge assumption transforms an adversary circuit $A$ into another (potentially polynomially larger) circuit $A'$ that behaves just like $A$ but also outputs an extracted value. If we apply a knowledge assumption recursively, then we would transform the original adversary circuit $A$ into $A'$, but then apply the knowledge assumption again to transform $A'$ into $A''$. However, clearly if such a recursion is applied a super-constant number of times, then the final circuit might be super-polynomial in size. This problem was encountered by Bitansky et al [BCCT12] in the construction of succinct non-interactive adaptive arguments of knowledge (SNARKs) using extractable collision resistant hash functions (ECRH). To prove the property of proof of knowledge, the extractor needs to extract the full Probabilistically Checkable Proof (PCP) given only the root of a Merkel tree. The natural solution is to apply the knowledge extraction recursively at each level of the tree. But since each level of extraction potentially incurs a polynomial blow up, one can apply extraction only a constant number of times. One of the major contributions of [BCCT12] was to circumvent this problem by using Merkel trees with polynomial fan-in and constant depth. Note that, however, we do not have any such option while constructing a constant round concurrent zero-knowledge protocol because the number of concurrent sessions can be any unbounded polynomial.

One natural approach to avoid this blow-up with each recursive extraction would be to assume a stronger property on the running time of the extractor. For example, one can assume the existence of an extractor which only takes an additive $poly(n)$ (where $n$ is the security parameter) factor more than the running time of the adversary. Note that the factor of $poly(n)$ is independent of the running time of the adversary. We call this the $+poly(n)$ assumption.

However, this assumption seems too strong and in fact potentially implausible[4]. On the other hand, if we do not make such a strong assumption, the essence of the problem is that if we want to apply the knowledge extractor recursively, we cannot afford it to take even $m^\epsilon$ longer than the adversary, where $m$ is the running time of the adversary and $\epsilon$ is an arbitrary constant. Note that we do not make the $+poly(n)$ assumption.

**Intuition behind our assumption.** Our first idea is to separate the process of extraction from the behavior of the adversary. More precisely, we will think of the adversary as a circuit $M$. If $M$ completes the Knowledge Commitment Protocol, an application of our assumption to $M$ gives us a *separate* extractor circuit $E$. The assumption states that the input wires of $E$ can be any wire inside the circuit $M$, including input, intermediate, or output wires. The output of $E(x)$ is *only* the value committed to in the output of $M(x)$. Now that we have separated the extractor from the adversary, we make the following observation: It is reasonable to assume that when the assumption is applied to create an extractor circuit $E$, the assumption does not attempt to place any "external knowledge" into $E$ or attempt to hide any knowledge in $E$. In other words, the extractor created by the assumption is *not* maliciously created. Hence, let us call it *benign* and denote it by $B$. Note that we will *only consider benign circuits that are created by the assumption*. The benign circuits are not assumed to remain benign if they are modified. Now we can state our assumption:

**Assumption 1.1** (Informal knowledge assumption)**.** Consider a pair of malicious and benign circuits $(M, B)$ such that $M$ completes a Knowledge Commitment Protocol and outputs a commitment to a value. Then there exists a polysize benign extractor circuit $E$ which takes as input a subset of wires of $M$, and outputs the value committed to by $M$. Moreover, the size of the extractor $E$ is bounded by a fixed polynomial in the size of $M$ and the security parameter $n$.

Now consider a recursive application of our assumption. Recall that the recursive application is required for the following: Suppose we have an adversary and we execute it to obtain some number of messages until it completes a Knowledge Commitment Protocol. Then we apply the knowledge assumption to obtain an extractor that allows us to obtain the committed value. We then use the extracted value in order to execute the adversary for some additional number of messages until it finishes another Knowledge Commitment Protocol (and so on). Let us denote by $M$ the execution of the adversary so far. Note that the inputs to $M$ are essentially the original inputs to the adversary together with the outputs of the extractors so far. Denote by $B$ the collection of extractors so far.

Now let us consider what happens when we apply our assumption to $(M, B)$. We obtain an extractor $E$ that extracts a value committed in the output of $M$. We observe that while $B$ was involved in the execution of the adversary, only the outputs of $B$ were ever used by $M$ to compute its output commitment message. Furthermore, as argued above, $B$ was benignly created by the assumption and thus has no external or hidden knowledge inside it. Thus we argue, that it is reasonable to assume that the size of the extractor $E$ created by the assumption is a fixed polynomial in the size of only the malicious circuit $M$. Recall that $M$ contains all the malicious computations done by the adversary. We now make the following observations about our assumption.

---

[4]Consider the following scenario: Given a random group element $g$ from a special group $G$, the adversary is expected to output $g^b$ (a commitment to $b$) and extractor's task is to output $b$. However, the Adversary applies a hash function on its input and gets a pseudorandom string $s = s_1 \ldots s_m$ of length $m$, where $m$ depends on the running time of the adversary and is not a fixed polynomial in length. Now, it traverses the string $s$ and recursively applies a special function $A$, such that $A(d, g^x) = g^{f(d,x)}$. In other words, the adversary computes $A(s_1, A(s_2, \ldots, A(s_m, g) \ldots))$. Now suppose $A$ and $f$ satisfy the following conditions: (1) $\mathsf{Time(A)} < \mathsf{Time(f)}$ (2) $\mathsf{Time}(f(s_1, f(s_2, \ldots, f(s_m, 1) \ldots))) = m \cdot \mathsf{Time(f)}$. Then, by the latter condition, the extractor needs to compute $f$ iteratively. Thus, the extractor will need at least $O(m)$ more operations than the adversary, where $m$ is decided by the adversary. We do not know if such an $A$ and $f$ exist. However, if such an $A$ and $f$ did exist, it would refute the $+poly(n)$ assumption.

- We observe that without loss of generality, we can assume that in a recursive application of our assumption, the extractor created by the assumption in fact contains all of the extractors created previously inside of it. Namely the benign circuit $B$ is a part[5] of the newly created extractor $E$. Thus $E$ can make use of all of the intermediate wires of previously created extractors, without loss of generality. These intermediate values may contain useful knowledge which may help the extraction of the value committed in the output message of $M$.

- We also observe that the counter-example we contemplated in Footnote 4 to the $+poly(n)$ assumption is compatible with our assumption[6]. That is, the existence of the functions $A$ and $f$ specified in the counter-example would not refute our assumption. Essentially this is because our extractor $E$ is *allowed to be polynomially larger* than the malicious circuit $M$.

- We further validate the plausibility of specific knowledge assumption that we make (see Section. 3.4) by providing a proof that the assumption holds in the generic group model (Section. 7).

- To understand what computational complexity limitations our assumption is placing on the Knowledge Commitment Protocol, let us first examine an important complexity limitation that the knowledge of exponent assumption of Hada and Tanaka (HTKEA) [HT98] places on the Knowledge Commitment Protocol. For simplicity of notation here, let us assume that the Knowledge Commitment Protocol is a non-interactive commitment denoted by $\mathsf{Com}(x)$. Consider a circuit $A$ such that:

$$A(x) = \overbrace{\mathsf{Com}(\mathsf{Com}\ldots(\mathsf{Com}(f(x)))\ldots)}^{\ell}$$

where $f$ is *not* efficiently computable. Then the HTKEA implies that there cannot exist such a polysize circuit $A$ for any constant $\ell$. This is because by making constant recursive invocations of HTKEA we will be able to extract $f(x)$ and generate a polysize circuit that computes $f$. Because our assumption admits further recursive invocations with efficient extractions, it would imply that such a polysize circuit $A$ should not exist for larger values of $\ell$. However, we note that the commitment we use is size increasing, namely $|\mathsf{Com}(x)| \geq 2|x|$. Therefore our assumption would imply that such a circuit $A$ cannot exist for any $\ell$ which is $O(\log(n))$. We believe that if such a complexity assumption holds for a constant $\ell$, as the HTKEA implies, then it is quite plausible that it holds for $\ell = O(log(n))$.

We describe two variants of our protocol: First, we provide a simpler protocol transformation that uses bilinear groups. This protocol is quite efficient and requires only 5 rounds. Our second protocol works with a knowledge assumption in general groups (without the need of a bilinear map), at the cost of a constant number of additional rounds, and is slightly less efficient.

**Organization.** The paper is organized as follows: We discuss the technical sections beginning with background on zero-knowledge, canonical arguments and commitment schemes in Section 2. We describe the *DHLA* assumption and our knowledge assumption for bilinear groups in Section 3. We describe our protocol (which uses bilinear groups) in Section 4 and prove its soundness in Section 5. Next, we show that our protocol is zero-knowledge in a concurrent setting in Section 6. For general groups, the knowledge assumption and the protocol for concurrent zero-knowledge is described in Appendix D. Then we prove that our knowledge assumption holds in the generic group model in Section 7. Finally, we discuss related work in Section 8.

---

[5] We stress that if all recursively created extractors contain all the previously created extractors inside it, then the last invocation of the assumption only needs to embed the previous extractor (since it already contains all previous extractors). This would prevent an exponential blow-up in size that a reader might otherwise worry would occur.

[6] On the other hand if the reader believes that the counter-example from Footnote 4 is not plausible, then it is easy to see that $+poly(n)$ assumption implies our assumption.

## 2 Definitions and Preliminaries

In the following sections, we will denote the security parameter by $n$. We denote a NP-complete language by $L$ and if $x \in L$ then $W_L(x)$ returns a witness $w$ to that fact.

**Definition 2.1** (Bilinear Groups). A bilinear group is a tuple $\mathcal{BG} = (q, \mathbb{G}, \mathbb{G}_T, e, g)$, where $\mathbb{G}$ and $\mathbb{G}_T$ are cyclic groups of prime order $q$, $g$ generates $\mathbb{G}$, and $e : \mathbb{G} \times \mathbb{G} \to \mathbb{G}_T$ is an efficient non-degenerate bilinear map, i.e. $\forall X, Y \in \mathbb{G} \; \forall a, b \in \mathbb{Z}_q : e(X^a, Y^b) = e(X, Y)^{ab}$, and $e(g, g)$ generates $\mathbb{G}_T$. Let $L_{QG}$ denote the set of $\{(q, g, e)\}$, where $g$ generates a bilinear group of prime order $q$, where $q$ is an $n$-bit prime, and $e$ is an efficient non-degenerate bilinear map. For brevity, we will suppress the bilinear map, when it is obvious from the context, and simply write $(q, g) \in L_{QG}$. Also, we will assume that if $q$ is an $n$−bit prime then any $x \in \mathbb{Z}_q$ can be represented by a unique $n$−bit string. For ease of notation, we just use $x$ to denote this unique string.

**Definition 2.2** (Interactive Arguments). Let $P, V$ be two PPT interactive machines. We denote the probability that $V$ accepts $x \in L$ on interacting with $P$ by $\mathsf{Acc}\langle P(x, w), V(x)\rangle$. We say that $\langle P, V \rangle$ is an interactive argument for an NP-complete language $L$ if the following two conditions are satisfied:

- Efficient Completeness: For every $x \in L$, there exists a string $w$, such that
$$\mathsf{Acc}\langle P(x, w), V(x)\rangle = 1.$$

- Computational Soundness: For every PPT machine $P^*$ (cheating prover), every polynomial $poly(\cdot)$, all sufficiently long $x \notin L$ and all strings $w$,
$$\mathsf{Acc}\langle P^*(x, w), V(x)\rangle < \tfrac{1}{poly(|x|)}.$$

**Definition 2.3** (Non-Black-Box Zero-Knowledge protocol w.r.t. auxiliary input of length $m$). Let $m$ be a polynomial in $n$. Let $P, V$ be two PPT interactive machines. We say that $\langle P, V \rangle$ is a non-black-box zero-knowledge protocol for $L$ w.r.t. auxiliary input of length $m$ if for every PPT machine $V^*$ there exists a PPT machine $S_{V^*}$ such that the following two distribution ensembles are indistinguishable:

$$\{S_{V^*}(x, y)\}_{x \in L, y \in \{0,1\}^m} \text{ and } \{\langle P(x, w), V^*(x, y)\rangle\}_{x \in L, w \in W_L(x), y \in \{0,1\}^m},$$

where $\{\langle P(x, w), V^*(x, y)\rangle\}_{x \in L, w \in W_L(x), y \in \{0,1\}^m}$ is a random variable taking the value of $V^*$'s random coins and the sequence of messages in the interaction between $P$ and $V^*$.

### 2.1 Concurrent Zero-Knowledge ($c\mathcal{ZK}$)

Let $\langle P, V \rangle$ be an interactive proof system for a language $L$, and consider a concurrent adversary $V^*$ that given an input $x \in L$ interacts with an unbounded number of copies of the prover $P$ concurrently. Moreover, there is no restriction on the scheduling of the messages between $P$ and $V^*$ (in particular, $V^*$ controls the scheduling of these messages).

The transcript of the concurrent session consists of the common input $x$, followed by a sequence of messages exchanged between the prover and the verifier. The view of $V^*$ when it interacts with $P$ consists of the random tape of $V^*$ together with the transcript of the protocol.

To prove that any protocol is zero-knowledge in the concurrent setting, we show the existence of a simulator for every concurrent verifier $V^*$ that interacts with $m$ copies of $P$, where $m$ is bounded by a polynomial in $n$.

**Definition 2.4** (Non-Black-Box $c\mathcal{ZK}$ with auxiliary input of length $m$). Let $\langle P, V \rangle$ be an interactive argument system for a language $L$. We say that $\langle P, V \rangle$ is non-black-box concurrent zero-knowledge if for every concurrent adversary $V^*$ (with auxiliary input $y$ of length $m$) that runs at most $m$ concurrent sessions with $P$, where $m$ is $n^c$ for any constant $c$, then there exists a probabilistic polynomial time algorithm $\mathcal{S}_{V^*}$ that runs in time polynomial in the running time of $V^*$ and $n$ and satisfies that the following ensembles are computationally indistinguishable:

$\{S_{m,V^*}(x,y)\}_{x \in L, y \in \{0,1\}^m, m \leq n^c}$ and $\{\langle P(x,w), V^*(x,y)\rangle\}_{x \in L, w \in W_L(x), y \in \{0,1\}^m, m \leq n^c}$

In the final constant round protocol for concurrent zero knowledge ($\Pi$) (see Section 4) using knowledge assumption in bilinear groups, we will use a discrete log based equivocal commitment scheme and three round canonical arguments as subroutines. Hence, we define and describe these next. Then we will describe the assumptions used to prove the soundness and the zero-knowledge properties of our protocol in bilinear groups. In the subsequent section, we will describe our protocol for concurrent zero-knowledge ($\Pi$) in detail. We describe the constant round concurrent zero-knowledge protocol for non-bilinear groups in Appendix D. In this protocol, we also use a constant round statistically sound zero-knowledge protocol in stand alone setting (see Appendix D.1).

## 2.2 Canonical Arguments

A three round canonical argument $\langle \overline{P}, \overline{V} \rangle$ for an NP-complete language $L$, proposed by [HT98], is described in Figure 1. CMT and RSP are the first and second messages of the prover and CH is the message sent by the verifier.

**Definition 2.5.** An argument system $\langle \overline{P}, \overline{V} \rangle$ for an NP-complete language $L$ is called a canonical argument system if it satisfies the following properties:

**B0** The prover is a probabilistic polynomial time function which is given the NP-witness $w$. When this function is invoked with an incoming message $M_{in}$ and its state, it outputs $M_{out}$ and its updated state. The initial state of the prover is set to $(x, w, R)$, where $x$ is the common input, $w$ is its auxiliary input and $R$ is the random tape. When it is invoked with $(\epsilon, (x, w, R))$ it outputs the prover's first message which is a commitment CMT.

**B1** The verifier selects the challenge CH uniformly at random from $\{0,1\}^n$.

**B2** Strong-Soundness: For any $x \notin L$ and CMT, there exists at most one challenge CH $\in \{0,1\}^n$ for which there exists a RSP $\in \{0,1\}^*$ such that $\text{VER}_x(\text{CMT}, \text{CH}, \text{RSP}) = 1$.

**B3** Honest Verifier Zero Knowledge (HVZK): There exists a probabilistic polynomial time Simulator $S_{HV}$ such that following two ensembles are computationally indistinguishable:

$$\{S_{HV}(x)\}_{x \in L} \text{ and } \{\langle \overline{P}(x,w), \overline{V}(x)\rangle\}_{x \in L, w \in W_L(x)},$$

where $\{\langle \overline{P}(x,w), \overline{V}(x)\rangle\}_{x \in L, w \in W_L(x)}$ is a random variable taking the value of $\overline{V}$'s internal coin tosses and the sequence of messages it receives in interaction between $\overline{P}$ (with auxiliary input $w$) and $\overline{V}$.

One of the ways to construct such a protocol, as described by Hada and Tanaka [HT98], is parallel composition of Blum's ZK protocol for Hamiltonicity.

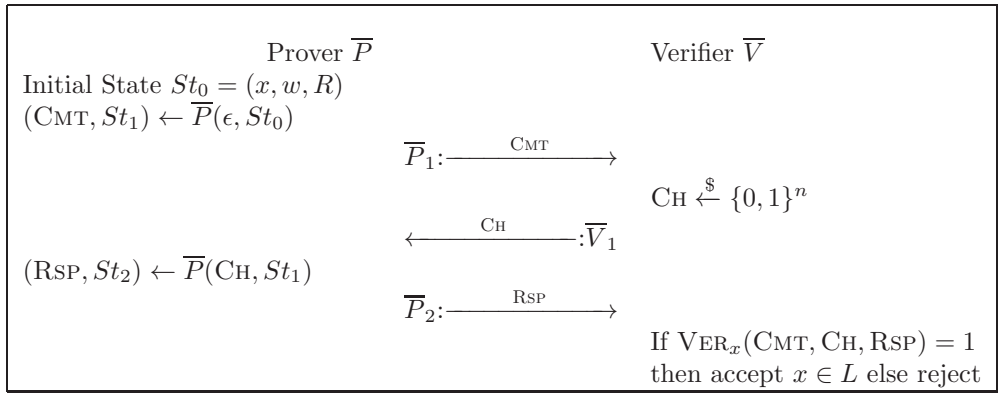

Figure 1: Three Round Canonical Argument System $\langle \overline{P}, \overline{V} \rangle$

## 2.3 Discrete Log based Equivocal Commitment Scheme $\mathsf{Com}_{DL}$

The committer and the receiver are given a group $\mathbb{G}$ of prime order $q$, its generator $g$ and an element $B \in \mathbb{G}$ such that $q$ is an $n$-bit prime. To commit to $x \in \mathbb{Z}_q$, choose $r \xleftarrow{\$} \mathbb{Z}_q$ and send $Z = g^x \cdot B^r$. To open, the sender sends $(x, r)$.

This commitment scheme is perfectly hiding i.e. $\mathsf{Com}_{DL}(x)$ and $\mathsf{Com}_{DL}(x')$ are identically distributed. If the committer does not know the discrete log of $B$, then $\mathsf{Com}_{DL}$ is computationally binding under discrete log assumption. We assume that discrete log assumption holds in all the groups we consider. Also, if $Z$ is a commitment under $\mathsf{Com}_{DL}$, then given two distinct openings of $Z$ to $(x, r)$ and $(x', r')$ such that $x \neq x'$, one can easily solve for the discrete log of $B$, say $b$, as follows: $b = (x - x') \cdot (r' - r)^{-1}$. Also, if the simulator knows the discrete log of $B$, say $b$, it can open $Z = \mathsf{Com}_{DL}(x; r)$ as being a commitment to any $x' \in \mathbb{Z}_q$ by sending $r' = \mathsf{Open}_{DL}(x, x', r, b) = (x + r \cdot b - x') \cdot b^{-1}$.

# 3 Assumptions

We begin this section by describing an assumption which is very similar to the discrete logarithm assumption (DLA). Given a $(q, g) \in L_{QG}$, DLA says that given a random group element $A = g^a$, for any polysize circuit, it is hard to compute $a$ with non negligible probability. *Diffie-Hellman Log Assumption* says that given a Diffie-Hellman tuple $(g^a, g^b, g^{ab})$, it is difficult to compute $b$ even when $a$ is chosen maliciously by the adversary. Let us denote Diffie-Hellman tuples by $\mathcal{DH}$.

**Assumption 3.1** (Diffie-Hellman Log Assumption (*DHLA*)). For every family of probabilistic polynomial size circuits $I = \{I_n\}_{n \geq 1}$, every $poly(\cdot)$, all sufficiently large $n$'s and all $(q, g) \in L_{QG}$ such that $q$ is of length $n$, consider the following probabilistic experiment:

- $I_n$ on input ( "Step 1", $1^n$) outputs $(g, A)$, where $A \in \mathbb{G}$.
- Given $(g, A)$ as input, experiment chooses $b \in \mathbb{Z}_q^*$ and computes $(B = g^b, X = A^b)$,

then *DHLA* says that if $(g, A, B, X)$ is a Diffie-Hellman tuple then the probability that $I_n$, given this tuple, outputs discrete log of $B$ is negligible even when $A$ is chosen maliciously by $I_n$. More formally,

$$Pr[I_n(\text{``Step 2''}, g, A, B, X | (A, B, X) \in \mathcal{DH}) = b : B = g^b] < \tfrac{1}{poly(n)},$$

for any choice of $A$ by $I_n$.

**Knowledge Assumption:** Below, by a circuit $C$ we mean a collection of Boolean gates and wires. We use the non-standard convention that certain gates are specially marked as output gates.

**Definition 3.2** (Admissible family of Adversaries). An admissible family of adversaries $\mathcal{A}$ is a family of sets such that the following properties hold: Each set $S \in \mathcal{A}$ is such that $S = \{C_n, M_n, B_n, \mathsf{aux}_n\}_{n \in \mathbb{N}}$. For each such set $S$, there exist constants $c, c' > 0$, such that $C_n$ is a circuit with $|C_n| \leq n^c$, and $\mathsf{aux} \subseteq \{0, 1\}^{n^{c'}}$. Furthermore, $\{M_n, B_n\}$ is a partition of the gates and the wires of the circuit $C_n$. If $x$ is the input to $C_n$ then by $M_n(x)$ we refer to the result of the computation $C_n(x)$ restricted to the output wires in $M_n$; we define $B_n(x)$ similarly.

We will refer to $M_n$ and $B_n$ as the malicious and the benign parts respectively of the adversary circuit $C_n$.

**Definition 3.3** ($\mathcal{A}$ *admits polysize malicious extensions*). An admissible family of adversaries $\mathcal{A}$ *admits polysize malicious extensions* if the following holds: For any set of circuits $S \in \mathcal{A}$ where $S = \{C_n, M_n, B_n, \mathsf{aux}_n\}_{n \in \mathbb{N}}$, and any polysize circuit family $\{F_n\}_{n \in \mathbb{N}}$ such that $\exists d > 0$, $|F_n| < n^d$ and the input wires to $F_n$ are a subset of the wires in $M_n$ (including both internal and output wires) and the output wires of $B_n$, we have that $S' = \{C_n \cup F_n, M_n \cup F_n, B_n, \mathsf{aux}_n\} \in \mathcal{A}$.

Next, based on the definition above, we define a variant of knowledge of exponent assumption based on the one described by Hada and Tanaka [HT98].

**Assumption 3.4.** [$m$-Knowledge of Exponent Assumption ($m$-KEA) w.r.t. admissible adversaries] We say that the $m$-Knowledge of Exponent Assumption holds with respect to a family of admissible adversaries $\mathcal{A}$, if for every $c > 0$, there exists a constant $c' > 0$ such that the following holds: For $m = n^c$, fix any $S = \{C_n, M_n, B_n, \mathsf{aux}_n\}_{n \in \mathbb{N}} \in \mathcal{A}$. Then there exists a family of extraction circuits $\{E_n\}_{n \in \mathbb{N}}$ whose inputs are a subset of any wires in $M_n$, such that $|E_n| \leq (n \cdot |M_n|)^{c'}$. (Informally, this condition requires that the extraction only uses the internal wires of the malicious part of the adversary.) Furthermore, we require that the following conditions hold:

1. For all sufficiently large $n$, every polynomial $poly(\cdot)$, the following is true for all $aux \in \mathsf{aux}_n$: Consider the following probabilistic experiment: For $i \in [1, m]$, primes $q_i$ and generators $g_i$ are chosen randomly such that $(q_i, g_i) \in L_{QG}$, where $q_i$ is chosen to be of length $n$. Values $a_1, \ldots, a_m$ are chosen at random such that $a_i \in \mathbb{Z}_{q_i}^*$. Finally, $R$ is chosen uniformly at random from sufficiently long strings so that the length of the tuple $x = ((q_1, g_1, g_1^{a_1}), \ldots, (q_m, g_m, g_m^{a_m}), aux, R)$ is exactly the length of the input to circuit $C_n$. If the input to $C_n$ is not long enough to allow such an input then the assumption is vacuously true for this $S$. Now, we consider the output of $M_n(x)$, which we interpret as a tuple $(j, B, X)$, where $j \in [m]$, and both $B$ and $X$ are in the group generated by $g_j$. Then, we interpret the output of $E_n(x)$ as the value $b_j \in \mathbb{Z}_{q_j}$, and require the following to be true:

$$Pr\left[X = B^{a_j} \wedge B \neq g_j^{b_j}\right] < \frac{1}{poly(n)}.$$

   (Informally, this condition states that if the malicious part of the adversary outputs a tuple so that $(g_j, g_j^{a_j}, B, X)$ form a Diffie-Hellman tuple, then the extractor $E_n$ successfully outputs the discrete log of $B$ with respect to $g_j$.)

2. We have that $(C_n \cup E_n, M_n, B_n \cup E_n, \mathsf{aux}_n) \in \mathcal{A}$. (Informally, this means that the extraction circuit created by this assumption is benign.)

**Definition 3.5.** An admissible set of adversaries $\mathcal{A}$ *contains all polysize malicious adversaries* if for all $c, c' > 0$, and for all circuit families $\{C_n\}_{n \in \mathbb{N}}$ such that $|C_n| \leq n^c$, for each $n$ there exists some subset $\mathsf{aux}_n \subseteq \{0, 1\}^{n^{c'}}$, such that $(C_n, C_n, \epsilon, \mathsf{aux}_n) \in \mathcal{A}$. We say that $\mathcal{A}$ *contains all polysize malicious adversaries with all polysize auxiliary inputs* if $\mathsf{aux}_n = \{0, 1\}^{n^{c'}}$ for each circuit family above.

**Theorem 3.6** (Informal)**.** *If the $m$-Knowledge of Exponent assumption holds with respect to an admissible adversary family $\mathcal{A}$ such that $\mathcal{A}$ contains all polysize malicious circuits and allow polysize malicious extension, and DHLA holds, then there exist constant-round concurrent zero-knowledge arguments for $\mathbf{NP}$ in the plain model.*

*Furthermore, if $\mathcal{A}$ contains all polysize malicious adversaries with all polysize auxiliary inputs, then there exist constant-round concurrent zero-knowledge arguments for $\mathbf{NP}$ in the plain model with respect to arbitrary auxiliary inputs.*

## 4 Constant Round Protocol for Concurrent Zero-Knowledge

The protocol starts by asking the verifier to use Knowledge Commitment Protocol to commit to a value $b$ in $B = g^b$. We use equivocal commitments whose trapdoor is $b$ to run a coin flipping protocol between the prover and the verifier. In parallel with the coin flipping protocol, we run a parallel repetition of Blum's Hamiltonicity protocol, where the result of coin flipping protocol determines the challenge message. We describe the 5-Round protocol for concurrent zero-knowledge argument in Figure 2. Note that the protocol execution does not make use of the bilinear map. It is only used by our zero-knowledge simulator to check that $(A, B, X)$ forms a Diffie-Hellman tuple since it does not have access to the discrete log of $A$. We stress that this

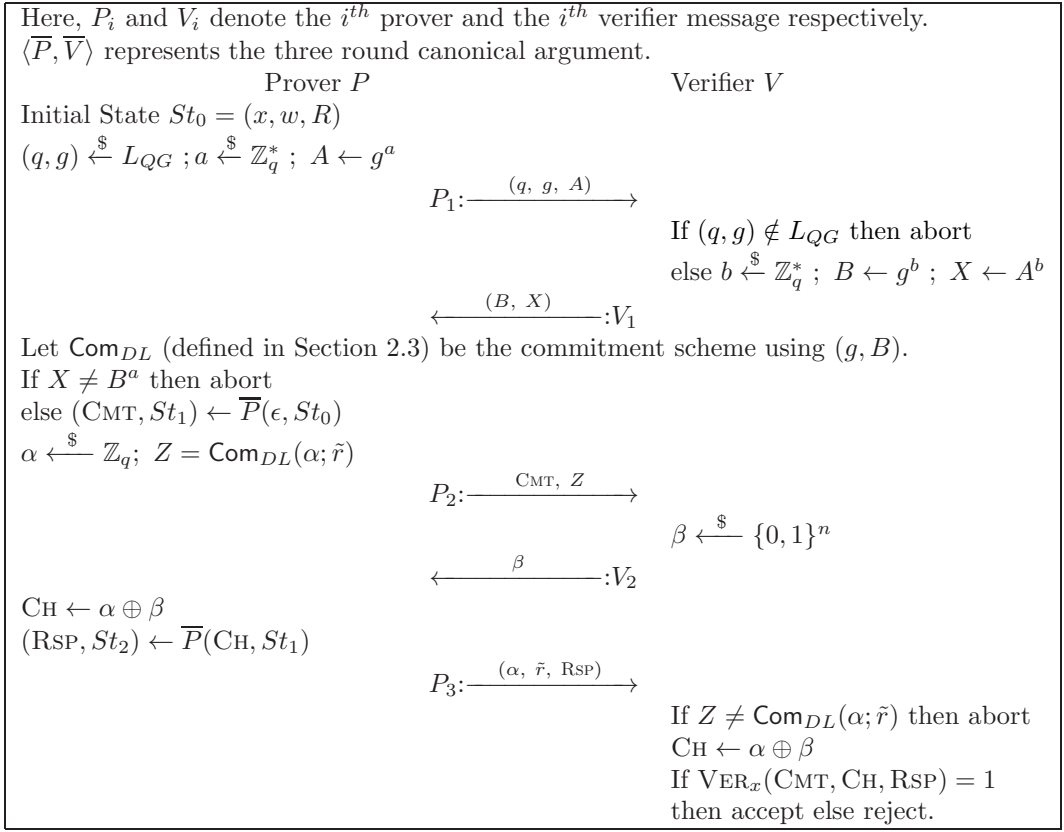

Here, $P_i$ and $V_i$ denote the $i^{th}$ prover and the $i^{th}$ verifier message respectively. $\langle \overline{P}, \overline{V} \rangle$ represents the three round canonical argument.

Prover $P$ — Verifier $V$

Initial State $St_0 = (x, w, R)$
$(q, g) \xleftarrow{\$} L_{QG}$ ; $a \xleftarrow{\$} \mathbb{Z}_q^*$ ; $A \leftarrow g^a$

$P_1: \xrightarrow{(q,\ g,\ A)}$

If $(q, g) \notin L_{QG}$ then abort
else $b \xleftarrow{\$} \mathbb{Z}_q^*$ ; $B \leftarrow g^b$ ; $X \leftarrow A^b$

$\xleftarrow{(B,\ X)} : V_1$

Let $\mathsf{Com}_{DL}$ (defined in Section 2.3) be the commitment scheme using $(g, B)$.
If $X \neq B^a$ then abort
else $(\mathrm{CMT}, St_1) \leftarrow \overline{P}(\epsilon, St_0)$
$\alpha \xleftarrow{\$} \mathbb{Z}_q$; $Z = \mathsf{Com}_{DL}(\alpha; \tilde{r})$

$P_2: \xrightarrow{\mathrm{CMT},\ Z}$

$\beta \xleftarrow{\$} \{0, 1\}^n$

$\xleftarrow{\beta} : V_2$

$\mathrm{CH} \leftarrow \alpha \oplus \beta$
$(\mathrm{RSP}, St_2) \leftarrow \overline{P}(\mathrm{CH}, St_1)$

$P_3: \xrightarrow{(\alpha,\ \tilde{r},\ \mathrm{RSP})}$

If $Z \neq \mathsf{Com}_{DL}(\alpha; \tilde{r})$ then abort
$\mathrm{CH} \leftarrow \alpha \oplus \beta$
If $\mathrm{VER}_x(\mathrm{CMT}, \mathrm{CH}, \mathrm{RSP}) = 1$
then accept else reject.

Figure 2: $\Pi$: 5-Round Protocol for $c\mathcal{ZK}$ $(P, V)$

use of a bilinear map is not crucial, and that we eliminate the need for a bilinear map in our second protocol (See Section D).

This protocol uses the discrete log based commitment scheme $\mathsf{Com}_{DL}$ which is binding under the hardness of *DHLA*. The secret value $b$ committed to by the verifier satisfies the following properties.

**R1:** For **Soundness:** Under *DHLA* (Assumption 3.1), any cheating prover while interacting with the honest verifier cannot get the secret coins of the verifier. Hence, any cheating prover cannot output the discrete log of $B$ sent by the verifier in Figure 2.

**R2:** For **Zero-knowledge:** Under *m-KEA* (Assumption 3.4), our simulator will be able to output the discrete log of $B$ no matter how the verifier behaves. Once the simulator gets the secret coins of $V^*$, which is the trapdoor to equivocal commitment scheme, the simulation is easy.

For **R2**, informally, it seems that even the cheating verifier must start by simply choosing $b$ and computing $(g^b, A^b)$ in order to pass the check $X = B^a$. That is, we assume that the verifier knows the secret coins $b$ whenever it passes the check. *m-KEA* defined in Section 3.4 captures this idea of knowledge and knowledge extraction formally. Under this variant of knowledge of exponent assumption, we will design a simulator which will extract the secret coins of the cheating verifier. Since, the simulator will have the trapdoor to $\mathsf{Com}_{DL}$, it will be able to equivocate on its commitment to $\alpha$ and force the outcome of the coin flipping protocol to the challenge string output by the honest verifier simulator $S_{HV}$.

# 5  Π is Computationally Sound

Recall that *DHLA* says that given a Diffie-Hellman tuple $(g, g^a, g^b, g^{ab})$, even if $a$ is chosen by the adversary, it is hard for it to guess $b$ with non-negligible probability. We prove soundness of Π by the following two steps: Let $P^*$ denote the cheating prover.

- If $P^*$ succeeds in equivocating its commitment in coin flipping protocol then we can extract the trapdoor value $b$ of Knowledge Commitment Protocol from $P^*$. This shows that $P^*$ can be used to efficiently compute $b$ and thereby break *DHLA*.

- We show that if $P^*$ does not equivocate on its commitment in coin flipping protocol and convinces the verifier of a false statement, then such a $P^*$ can be used to violate the underlying strong soundness of canonical arguments. In other words, it would violate the underlying soundness of Blum's Hamiltonicity protocol.

To prove the soundness of Π in the concurrent setting, it is sufficient to prove soundness of a single stand alone session.

For the first step, let us define the following interactive game $\mathcal{G}$:

1. $Sim$ runs the above protocol with $P^*$ till $P^*$ commits to $\alpha$ using random coins $\tilde{r}$ in the above protocol using commitment scheme $\mathsf{Com}_{DL}$ as defined before. $P^*$ sets $Z = \mathsf{Com}_{DL}(\alpha; \tilde{r})$ and sends $Z$ to $Sim$.

2. $Sim$ sends $\beta$ to $P^*$.

3. $P^*$ sends $(\alpha_1, \tilde{r}_1, \text{RSP})$ to $Sim$ such that $Z = \mathsf{Com}_{DL}(\alpha_1; \tilde{r}_1)$.

4. $Sim$ rewinds $P^*$ to Step 2 and sends it $\beta'$ such that $\beta' \neq \beta$. $P^*$ wins if it sends $(\alpha_2, \tilde{r}_2, \text{RSP})$ to $Sim$ such that $Z = \mathsf{Com}_{DL}(\alpha_2; \tilde{r}_2)$ and $\alpha_1 \neq \alpha_2$.

**Lemma 5.1.** Under *DHLA*, for every probabilistic polynomial time machine $P^*$, every polynomial $poly(\cdot)$, and all sufficiently large $n$'s,

$$Pr[P^* \text{ wins } \mathcal{G}] < \frac{1}{poly(n)}$$

where probability is over choice of $\alpha, \beta$ and coins of $P^*$ and $n$ is the security parameter.

**Proof:** We will prove this by contradiction. If there is a polynomial $f(n)$ such that $Pr[P^* wins\ \mathcal{G}] > 1/f(n)$, then we can construct an adversary $\mathcal{A}$ for *DHLA*. $\mathcal{A}$ runs $P^*$ and gets $(q, g, A)$ and sends $(g, A)$ to the challenger $\mathcal{Ch}$ of *DHLA*. $\mathcal{Ch}$ prepares the challenge tuple by choosing a random $b$ and sends $(B = g^b, X = A^b)$ to $\mathcal{A}$ which it forwards to $P^*$. $P^*$ and $\mathcal{A}$ continue running the protocol Π until the opening of $Z$ as $\alpha$. After this opening, $\mathcal{A}$ rewinds $P^*$ until the commitment $Z$ and runs $P^*$ again with a different $\beta'$ and looks at the opening of $Z$ by $P^*$. If $P^*$ opens $Z$ to the same $\alpha$, $\mathcal{A}$ aborts. Else if $P^*$ opens $Z$ to a $\alpha'$ such that $\alpha \neq \alpha'$, $\mathcal{A}$ can compute $b$, the discrete log of $B$, as described in Section 2.3. $\mathcal{A}$ sends $b$ to $\mathcal{Ch}$.
$Pr[\mathcal{A} \text{ breaks } DHLA] = Pr[P^* \text{ wins } \mathcal{G}] > 1/f(n)$. This contradicts *DHLA*.

**Theorem 5.2.** *Under Lemma 5.1 and strong soundness property (B2) of $\langle \overline{P}, \overline{V} \rangle$, protocol Π is computationally sound.*

**Proof:** We will again prove this by contradiction. If Π is not computationally sound then there exists a PPT machine $P^*$ and an infinite set $\mathcal{I} = \{(x, w) : x \notin L\}$ such that there exists a polynomial $p(\cdot)$ satisfying

$$\mathsf{Acc}\langle P^*(x, w), V(x) \rangle > \frac{1}{p(|x|)}.$$

Since $P^*$ can equivocate his commitment to $\alpha$ only with a negligible probability, the only way $P^*$ can convince $V$ of a false statement is to complete the protocol successfully for multiple challenges for each $(x, w) \in \mathcal{I}$.

Using this cheating prover $P^*$ for Π, we will construct a cheating prover $\overline{P}^*$ for canonical argument system which breaks the strong soundness property of the canonical argument system. $\overline{P}^*$ interacts with $P^*$ as the verifier on some $(x, w) \in \mathcal{I}$. $\overline{P}^*$ runs the protocol till the end. If $P^*$ succeeds in convincing $\overline{P}^*$, then $\overline{P}^*$ rewinds $P^*$ up to the point when $P^*$ has sent his commitment to $\alpha$. This time $\overline{P}^*$ sends a different $\beta$. If $P^*$ completes the protocol successfully for the second time, then $\overline{P}^*$ gets two different tuples $(\text{CMT}, \text{CH}_1, \text{RSP}_1)$ and $(\text{CMT}, \text{CH}_2, \text{RSP}_2)$ for $x \notin L$ which a honest verifier $\overline{V}$ would accept. This is because $P^*$ can not equivocate to $\alpha$ with non-negligible probability. This would contradict the strong soundness property of $\langle \overline{P}, \overline{V} \rangle$.

Now we need to calculate the probability of success of $\overline{P}^*$ in breaking **B2**.

Let $\mathcal{T}$ be the transcript of $\Pi$ till the prover's commitment to $\alpha$. Let $p_\mathcal{T} = Pr[V \text{ accepts } x \notin L | \mathcal{T}]$. We are given that,

$$\text{Acc}\langle P^*(x, w), V(x) \rangle = \mathbb{E}_\mathcal{T}(p_\mathcal{T}) > \tfrac{1}{p(|x|)}.$$

$$\tfrac{1}{p(|x|)} < \mathbb{E}_\mathcal{T}(p_\mathcal{T}) \leq \left( Pr_\mathcal{T} \left[ p_\mathcal{T} > \tfrac{1}{2} \cdot \tfrac{1}{p(|x|)} \right] \cdot 1 \right) + \left( Pr_\mathcal{T} \left[ p_\mathcal{T} \leq \tfrac{1}{2} \cdot \tfrac{1}{p(|x|)} \right] \cdot \tfrac{1}{2} \cdot \tfrac{1}{p(|x|)} \right).$$

Now we know that $Pr_\mathcal{T} \left[ p_\mathcal{T} \leq \tfrac{1}{2} \cdot \tfrac{1}{p(|x|)} \right] \leq 1$. By solving we get,

$$Pr_\mathcal{T}[p_\mathcal{T} > \tfrac{1}{2} \cdot \tfrac{1}{p(|x|)}] > \tfrac{1}{2p(|x|)}.$$

We call a transcript $\mathcal{T}$ good when $p_\mathcal{T} > \tfrac{1}{2} \cdot \tfrac{1}{p(|x|)}$. $\overline{P}^*$ breaks **B2** on a good transcript when $\overline{P}^*$ succeeds in convincing the verifier on two independent choices of $\beta$. Hence,

$$Pr[\overline{P}^* \text{ breaks } \mathbf{B2}] > Pr[\mathcal{T} \text{ is good}] \cdot (\tfrac{1}{2} \cdot \tfrac{1}{p(|x|)})^2 > \tfrac{1}{8} \cdot (\tfrac{1}{p(|x|)})^3.$$

This contradicts the strong soundness property of canonical argument system by a non-negligible probability.

## 6 $\Pi$ is Concurrently Zero-Knowledge

To establish the zero-knowledge property, we build a sequence of extractors through recursive applications of *m-KEA*. Informally, each circuit uses the extractor provided by *m-KEA* to obtain the value $b$ committed by $V^*$ and then use this trapdoor value to equivocate in the coin flipping protocol. Through such an equivocation, the simulator can force the challenge message in Blum's Hamiltonicity protocol to be equal to the challenge the simulator received by calling the honest verifier simulator $S_{HV}$ for Blum's Hamiltonicity. We prove that the simulation is computationally indistinguishable from the real execution through a sequence of hybrids.

**Theorem 6.1.** *If there are $m$ concurrent sessions of $\Pi$ and if our family of admissible adversaries $\mathcal{A}$ contains all polynomial size adversaries and allows polysize malicious extensions, then under m-KEA and honest verifier zero-knowledge property of $\langle \overline{P}, \overline{V} \rangle$, the following distribution ensembles are computationally indistinguishable:*

$$\{S_{V^*}(x, y)\}_{m, x \in L, y \in \{0,1\}^m} \text{ and } \{\langle P(x, w), V^*(x, y) \rangle\}_{m, x \in L, w \in W_L(x), y \in \{0,1\}^m},$$

*where $\mathcal{S}_{V^*}$ is the zero-knowledge simulator for $\Pi$ described in Appendix B.*

For the proof of this theorem refer to Appendix B. The following theorem states that the circuit of our simulator $\mathcal{S}_{V^*}$ is a polynomial size circuit.

**Theorem 6.2.** *The size of the circuit of the simulator $\mathcal{S}_{V^*}$ is a fixed polynomial in the size of the circuit of $V^*$ and the security parameter.*

For the proof of this theorem refer to Appendix C.

## 7 *m-KEA* holds in Bilinear Generic Group Model

In this section, we will argue that *m-KEA* (Assumption 3.4) holds for any family of admissible adversaries (described below) that acts generically to the groups used in our protocol.

**The Generic Group model:** Given a cyclic bilinear group $\mathbb{G}$, we consider the random encoding $\psi_G$, that is an injective map $\psi_G : \mathbb{G} \to \{0, 1\}^\ell$, where $\ell > 3 \cdot \log(|\mathbb{G}|)$. We write the encoded group as $\{\psi_G(x) : x \in \mathbb{G}\}$. Let $e$ be the bilinear map, $e : \mathbb{G} \times \mathbb{G} \to \mathbb{G}_T$, where $\mathbb{G}_T$ is also a cyclic group. Let the random encoding of $\mathbb{G}_T$ be $\psi_{G_T}$. The adversary is given access to three oracles $\Phi^B$, $\Phi^P$, and $\Phi^T$. The oracle $\Phi^B$ takes as input the random encodings of group elements in base group $\mathbb{G}$ and performs the group operations multiplication and inverse in $\mathbb{G}$. If $\alpha, \beta \in \psi_G(\mathbb{G})$, then $\Phi^B(\alpha, \beta)$ gives the encoding of the product of elements represented by $\alpha$ and $\beta$ which is $\psi_G(\psi_G^{-1}(\alpha) \cdot \psi_G^{-1}(\beta))$. Also $\Phi^B(\alpha, \mathsf{Inv})$ gives the encoding of the inverse of the group element represented by $\alpha$ which is $\psi_G((\psi_G^{-1}(\alpha))^{-1})$, where $\alpha \in \psi_G(\mathbb{G})$. $\Phi^P$ also takes random encodings of group elements in $\mathbb{G}$ and returns the pairing under the bilinear map $e$. Given $\alpha, \beta \in \psi_G(G)$, $\Phi^P(\alpha, \beta)$ gives the encoding of the pairing of elements represented by $\alpha$ and $\beta$ which is $\psi_{G_T}(e(\psi_G^{-1}(\alpha), \psi_G^{-1}(\beta)))$. Similarly, $\Phi^T$ takes as input the random encodings of group elements in $\mathbb{G}_T$ and performs the group operations multiplication and inverse in $\mathbb{G}_T$. If $\alpha, \beta \in \psi_{G_T}(\mathbb{G}_T)$, then $\Phi^T(\alpha, \beta)$ gives the encoding of the product of elements represented by $\alpha$ and $\beta$ which is $\psi_{G_T}(\psi_{G_T}^{-1}(\alpha) \cdot \psi_{G_T}^{-1}(\beta))$. Also $\Phi^T(\alpha, \mathsf{Inv})$ gives the encoding of the inverse of the group element represented by $\alpha$ which is $\psi_{G_T}((\psi_{G_T}^{-1}(\alpha))^{-1})$, where $\alpha \in \psi_{G_T}(\mathbb{G}_T)$. Let $\Phi = \Phi^B \cup \Phi^P \cup \Phi^T$. Without loss of generality, assume $\psi_G(1_G) = 0^\ell$.

**Theorem 7.1.** *m-KEA holds in the bilinear generic group model w.r.t. a family of admissible adversaries $\mathcal{A}$ which contains all polysize malicious circuits with all polysize auxiliary inputs.*

**Proof:** Consider the following family of admissible adversaries $\mathcal{A} = \{(C_n, M_n^\Phi, B_n, \mathsf{aux}_n)\}_{n \in \mathbb{N}}$, where $M_n^\Phi$ is any family of polysize malicious circuits which have access to the oracle $\Phi$, i.e. $\Phi^B$, $\Phi^P$ and $\Phi^T$. $B_n$ is any family of polysize circuits which do not make any calls to any of the oracles. $C_n = M_n^\Phi \cup B_n$ and $\mathsf{aux}_n$ is the set of all polysize strings. Observe that $\mathcal{A}$ admits any polysize malicious extension. Looking ahead, the family of extractor circuits $E_n$ will not make any calls to $\Phi$.

Since in the experiment described in m-KEA, the circuit $C_n$ deals with $m$ different bilinear groups, the oracles $\Phi^B$, $\Phi^P$ and $\Phi^T$ also take the group number $i$ as input. More formally, if $\alpha, \beta \in \psi_{G_i}(\mathbb{G}_i)$, then $\Phi^B(i, \alpha, \beta) = \psi_{G_i}(\psi_{G_i}^{-1}(\alpha) \cdot \psi_{G_i}^{-1}(\beta))$, $\Phi^B(i, \alpha, \mathsf{Inv}) = \psi_{G_i}((\psi_{G_i}^{-1}(\alpha))^{-1})$ and $\Phi^P(i, \alpha, \beta) = \psi_{G_{T,i}}(e_i(\psi_{G_i}^{-1}(\alpha), \psi_{G_i}^{-1}(\beta)))$. Similar calls can be made to $\Phi^T$ to compute the multiplication and inverse operations in the groups $\mathbb{G}_{T_i}$.

To prove the second property of m-KEA, we would maintain the invariant that our extractor circuit family $E_n$ will not make any oracle calls. Now, in order to prove the theorem, we are left to prove that the first property of m-KEA holds.

For all sufficiently large $n$, given a set of circuits $S = \{C_n, M_n^\Phi, B_n, \mathsf{aux}_n\} \in \mathcal{A}$ (defined above) and for all $aux \in \mathsf{aux}_n$, we run the following experiment. For all $i \in [m]$, we pick at random $(q_i, g_i) \in L_{QG}$ and pick a random encoding $\psi_{G_i}$ for the group $\mathbb{G}_i$ and $\psi_{G_{T_i}}$ for the group $\mathbb{G}_{T_i}$. Let $\Phi$ be the oracle for the bilinear generic group model. Now, choose values $a_1, a_2, \ldots, a_m$ uniformly at random such that $a_i \in \mathbb{Z}_{q_i}^*$. The circuit $M_n^\Phi$ is given the input $(q_i, \psi_{G_i}(g_i), \psi_{G_i}(g_i^{a_i}))$ for all $i \in [m]$. Let the output of $M_n^\Phi$ be $(j, B, X)$. Now we construct the extractor circuit $E_n$ as follows:

$E_n$ builds a lookup table $T_i$ for all $i \in [m]$. Table $T_i$ maps all group elements ($\alpha = \psi_{G_i}(g_i^{d_i + a_i \cdot d_i'})$) ever considered by $M_n^\Phi$ to a tuple $((d_i, d_i'))$. Informally, it maps the encoding of any group element to its discrete log w.r.t. $g_i$ and $g_i^{a_i}$. Specifically, $E_n$ works as follows: It initializes $T_i$ with following two entries: $\psi_{G_i}(g_i)$ maps to $(1, 0)$ and $\psi_{G_i}(g_i^{a_i})$ maps to $(0, 1)$.

Now, consider a topological ordering $\tau$ of the gates in circuit $M_n^\Phi$. Traversing in this order, we process the oracle calls to $\Phi^B$ as follows: For oracle call $\Phi^B(i, \alpha, \beta)$ with outcome $\gamma \neq \bot$, we update our $T_i$ as follows: Find the entries in $T_i$ corresponding to $\alpha$ and $\beta$, say $T_i(\alpha)$ and $T_i(\beta)$. If any one of these is not found, we call this event Miss and $E_n$ outputs MissFail. Then find $\gamma$ in $T_i$. If $T_i(\gamma)$ exists, but $T_i(\gamma) \neq T_i(\alpha) + T_i(\beta)$ (addition is done component wise and modulo $q_i$), we call this event Collision and $E_n$ outputs CollisionFail. Else, set $T_i(\gamma) = T_i(\alpha) + T_i(\beta)$. For the oracle call $\Phi^B(i, \alpha, \mathsf{Inv})$ with outcome $\gamma \neq \bot$, we update our $T_i$ as follows: Find the entry corresponding to $\alpha$. If it does not exist, we call this event Miss and $E_n$ outputs MissFail. If $T_i(\gamma)$ already exists but $T_i(\gamma) \neq -T_i(\alpha)$, then this event is Collision and $E_n$ outputs CollisionFail. Else, set $T_i(\gamma) = -T_i(\alpha)$. For the oracle calls to $\Phi^P$ and $\Phi^T$, $E_n$ does nothing.

After processing all the gates in $M_n^\Phi$ which make an oracle call, $E_n$ looks at the output wires of $M_n^\Phi$ and interprets it as a tuple of the form $(j, B, X)$ for some $j \in [m]$. If $T_j(B)$ or $T_j(X)$ does not exist, we call this event OutMiss and output OutFail. Else if, $T_j(B) = (b, 0)$ and $T_j(X) = (0, b)$, $E_n$ outputs $b$ else $E_n$ outputs Fail. Consider the event when $(\psi_{G_j}^{-1}(B))^{a_j} = \psi_{G_j}^{-1}(X)$ but $T_j(B)$ and $T_j(X)$ are not of the form described above. We call this event FalseNegative because though $M_n^\Phi$ outputs a valid tuple, but $E_n$ fails to output the discrete log.

Observe that as stated earlier, $E_n$ does not make any oracle calls. Instead, it only examines the inputs and outputs of oracle calls made by $M_n^\Phi$. Since $M_n^\Phi$ can make at most $|M_n^\Phi|$ oracle calls, the total size of all the lookup tables is at most $|M_n^\Phi|$. Hence, $\exists c > 0$ such that $|E| \leq (n \cdot |M_n^\Phi|)^c$. There are four cases in which $E_n$ misbehaves and below we prove that the probability of each of these events is negligible.

1. Event Miss: This event happens when $M_n^\Phi$ makes an oracle call with input which is neither an input to $M_n^\Phi$ nor an output of some previous oracle call. This happens when $M_n^\Phi$ is able to guess an $\ell - bit$ string which is a valid encoding of some group element. Hence, $\Pr[\mathsf{Miss}] \leq \max_i \frac{|M_n^\Phi| \cdot q_i}{2^{\ell_i}} \leq \max_i \frac{|M_n^\Phi|}{q_i^2}$, which is negligible since $M_n^\Phi$ is polysize.

2. Event Collision: This happens when there is a conflict between old $T_i(\gamma)$ (say,$(x_1, y_1)$) and new output from $\Phi^B$ for $T_i(\gamma)$ (say, $(x_2, y_2)$). This would give us an equation of the form $x_1 + y_1 \cdot a_i = x_2 + y_2 \cdot a_i$, where the only unknown is $a_i$. Solving this equation, $E_n$ can learn the value of $a_i$. But $a_i$ was information theoretically hidden. So, $\Pr[\mathsf{Collision}]$ for any oracle call is at most $\frac{1}{q_i}$. Taking union bound over all the oracle calls in $M_n^\Phi$, $\Pr[\mathsf{Collision}] \leq \max_i \frac{|M_n^\Phi|}{q_i}$. Since, $M_n^\Phi$ is polysize, this probability is negligible.

3. Event FalseNegative: This event happens when $(\psi_{G_j}^{-1}(B))^{a_j} = \psi_{G_j}^{-1}(X)$ but $T_j(B)$ and $T_j(X)$ are not of the form $(b, 0)$ and $(0, b)$ respectively. Let $T_j(B) = (x_1, y_1)$ and $T_j(X) = (x_2, y_2)$. Note that since $(\psi_{G_j}^{-1}(B))^{a_j} = \psi_{G_j}^{-1}(X)$ it is the case that $a_j \cdot (x_1 + y_1 \cdot a_j) = x_2 + y_2 \cdot a_j$. This gives a quadratic equation in $a_j$ which has at most two roots. Hence, $\Pr[\mathsf{FalseNegative}] \leq \max_i \frac{2}{q_i}$.

4. Event OutMiss: We are only concerned in the event when $E_n$ outputs OutFail but $(B, X)$ is a valid tuple of group elements. This happens when $M_n^\Phi$ successfully guesses at least one $\ell-$bit string which is a valid encoding of some group element. As argued above for the event Miss, this probability is bounded above by $\max_i \frac{|M_n^\Phi|}{q_i^2}$. This bound must also hold for the event OutMiss. Hence $\Pr[\mathsf{OutMiss} \wedge \mathsf{Valid}] \leq \max_i \frac{|M_n^\Phi|}{q_i^2}$, which is negligible.

We have shown the construction of $E_n$ such that if $M_n^\Phi$ outputs a tuple $(j, B, X)$ such that $(\psi_{G_j}^{-1}(B))^{a_j} = \psi_{G_j}^{-1}(X)$, then $E_n$ outputs $b$ such that $\psi_{G_j}^{-1}(B) = g_j^b$ with all but negligible probability. Since $E_n$ does not make any oracle calls, $\{C_n \cup E_n, M_n^\Phi, B_n \cup E_n, \mathsf{aux}_n\} \in \mathcal{A}$. Hence, m-KEA assumption holds for $\mathcal{A}$ defined above.

*Remark:* Note that because the adversary has to output the tuple $(B, X)$ in the base group $\mathbb{G}$, the calls to the oracles $\Phi^P$ and $\Phi^T$ are simply irrelevant to the proof. Hence, they neither arise in construction of the extractor $E_n$ nor cause any complication to its existence. Moreover, almost the same extractor construction can be used to show that knowledge assumption for non-bilinear groups (see Appendix D) would hold in generic groups.

## 8 Related Work

**Knowledge Assumptions** Knowledge or extractability assumptions capture our belief that certain computational tasks can be done efficiently only by going through certain specific intermediate stages and generating some specific kinds of intermediate values. One such class of assumptions is Knowledge of Exponent Assumptions which were first introduced by Damgard [Dam91] to construct a CCA secure encryption scheme. Though these assumptions do not fall in the class of falsifiable class of assumptions [Nao03], these have been proven secure against generic algorithms [Nec94, Sho97, Den06], thus offering some evidence of validity. Hada and Tanaka [HT98] gave a three round zero-knowledge protocol using two knowledge of exponent assumptions. Later, Bellare and Pallacio [BP04] proved that the assumption used for proving the soundness of the protocol was false, proposed a modified assumption and recovered the earlier result. We stress that in our protocol, we are able to argue soundness directly without the use of any knowledge assumption.

Extending the assumption of [BP04], Abe and Fehr [AF07] constructed the first perfect NIZK for NP with full adaptive soundness. Under knowledge of exponents assumption, Prabharakaran and Xue [PX09] constructed statistically hiding sets based on trapdoor DDH groups [DG06]. Gennaro et al. [GKR10] modify the Okamoto-Tanaka key agreement protocol to get perfect forward secrecy. Recently, Groth [Gro10] generalized the assumption of [AF07] to short non-interactive perfect zero-knowledge arguments for circuit satisfiability.

Other set of knowledge assumptions used recently are extractable functions [CD08, CD09]. All of [BCCT12, DFH12, GLR11] give one of the constructions of Extractable Collision Resistant Hash functions (ECRH) using Knowledge of Exponent Assumptions. Then assuming the existence of ECRH, Bitansky et al [BCCT12] modify the construction of [CL08] and prove that the modified construction is a succinct non-interactive adaptive arguments of knowledge (SNARK). They also show that existence of SNARKs imply the existence of (their notion of) ECRH. In the CRS model, they combined NIZK and SNARKs to give zero-knowledge non-interactive arguments. On the other hand, Damgard et al [DFH12] also use ECRH to construct succinct non-interactive arguments in CRS model. Using these, they give a two message protocol for two party computation which is UC-secure.

**Concurrent Zero-Knowledge:** The difficulty in constructing a round-efficient $c\mathcal{ZK}$ was first observed by Dwork et al. [DNS98]. Following this, rigorous lower bounds on round complexity of $c\mathcal{ZK}$ for NP with a black-box simulator have been proven in [KPR98, Ros00, CKPR01]; the best lower bound being $\Omega(\log n / \log \log n)$ rounds given by Canetti et al. [CKPR01].
Barak [Bar01] gave a constant round protocol for all NP, in which he gave a non-black-box simulator for zero-knowledge. Also, for any predetermined polynomial $p(\cdot)$, this constant round protocol is zero-knowledge even when $p(n)$ sessions are concurrently executed. But it has a major drawback. The polynomial $p(\cdot)$ has to be fixed at the beginning of the protocol and the message lengths grow linearly in $p(n)$. Killian and Petrank [KP01] gave a poly-logarithmic round protocol which is zero-knowledge even when it is executed concurrently for any (not determined) polynomial number of times. The gap between the upper and lower bounds of round complexity of black-box $c\mathcal{ZK}$ was closed by Prabhakaran, Rossen, and Sahai [PRS02] who gave a $\tilde{O}(\log n)$ round protocol. Since then improving the round complexity of concurrent zero-knowledge has been an open problem.

# A Discussion regarding use of auxiliary inputs for concurrent simulation

A potentially promising idea for using knowledge assumptions for concurrent simulation is the following: Formulate a knowledge assumption that holds for all auxiliary inputs for the adversary, and then invoke the knowledge extractor provided by the knowledge assumption with different auxiliary inputs corresponding to the extraction history. In other words, one could attempt to apply a single extractor iteratively for different concurrent sessions, passing along all the information extracted so far as auxiliary input to the extractor.

However, similar to the example discussed in the Introduction concerning a potential "interactive" knowledge assumption, a problem may arise if the auxiliary input contains "external knowledge" and thereby prevents extraction. We stress there is an important distinction between why this fails and failure of the interactive knowledge assumption. Here we are not saying that a knowledge assumption which holds with regard to all auxiliary inputs must be false. Rather we are saying that any natural application of such an assumption to the concurrent setting would fail. This is because it would cause us to invoke the extractor with auxiliary inputs that impermissibly correlate with messages received by the adversary in earlier executions of Knowledge Commitment Protocol. By the definition of auxiliary input, an extractor would not be required to function in such a case. To make the intuition precise, consider an example of such an iterative application of knowledge assumption in the concurrent setting. Suppose the Adversary schedules the messages of the malicious committer (MC) as follows: First, MC asks for the random first message of the Receiver (R) in the Knowledge Commitment Protocol for all the sessions $(r_1, r_2, \ldots, r_m)$. Now, MC chooses a function $f$ and completes the first Knowledge Commitment Protocol by committing to $f(r_1, r_2, \ldots, r_m)$. We apply the knowledge assumption to recover $f(r_1, r_2, \ldots, r_m)$. Next, the MC completes another Knowledge Commitment Protocol. Now in order to extract, we need to provide the extractor one of the random $r_i$'s as input and $f(r_1, r_2, \ldots, r_m)$ as auxiliary input. But here, depending on the function $f$, this auxiliary input may be highly correlated to the input $r_i$. In this case, the extractor is *allowed* to fail with high probability. This is because the extractor is only required to work for the fixed auxiliary input $aux = f(r_1, r_2, \ldots, r_m)$, when $r_i$ is chosen at random independently of $aux$. However, the actual simulation would use $aux$ that correlates with the input $r_i$.

## B  Description of the simulator

In the concurrent setting, the verifier may start an unbounded number of sessions with the prover and may interleave them in any way he wants. One such individual session has five rounds (as shown in Figure 2). In this section, we will model our cheating verifier $V^*$ as a next message function with a state $\gamma$.

$$V^*(Msg', k, \gamma') \to (Msg, j, \gamma, \ell)$$

where $Msg'$ is the prover's (or simulator's) message from the session $k$ and $\gamma'$ is the last state of $V^*$. In response, $V^*$ sends message $Msg$ corresponding to some session $j$ and changes its state to $\gamma$. Prover's (or simulator's) next message would be the next message from the session $j$. In case $Msg$ is $\epsilon$, then the verifier is requesting for the first message of session $\ell$. Verifier can also output a special message (END, output), which means that $V^*$ wants to stop the execution with output output.

To describe our simulator $\mathcal{S}_{V^*}$, we will first describe a sequence of admissible adversaries $\{C_{n,i}, M_{n,i}, B_{n,i}, \mathsf{aux}_n\}$ and $\{C'_{n,i}, M'_{n,i}, B'_{n,i}, \mathsf{aux}_n\}$ for all $i \in \{1, 2, \ldots, m+1\}$. First, we will describe these for $i = 1$ followed by $i > 1$ recursively using $\{C_{n,i-1}, M_{n,i-1}, B_{n,i-1}, \mathsf{aux}_n\}$ and $\{C'_{n,i-1}, M'_{n,i-1}, B'_{n,i-1}, \mathsf{aux}_n\}$. Each of these circuits will maintain and update the set of aborted sessions called Aborted. We will assume that the simulator knows the upper bound on $m$, the number of sessions that $V^*$ executes. Also, whenever $V^*$ stops, our simulator stops with the output of $V^*$.

**Admissible adversary:** $\{C_{n,1}, M_{n,1}, B_{n,1}, \mathsf{aux}_n\}$.
**Input:** $(x, y, (q_1, g_1, g_1^{a_1}), \ldots, (q_m, g_m, g_m^{a_m}))$ and $(R_1, R_2)$, where $x \in L$, $y$ is the auxiliary input of length $m$ and $(q_i, g_i) \in L_{QG}$, for all $i$. $R_1$ is the random tape for $C_{n,1}$ and $R_2$ is the random tape for $V^*$.
**Output:** $(j, B_j, X_j)$ or (END, output).
**Description:** We will start building the circuit $F_{n,1}$ as follows: $F_{n,1}$ will simulate the interaction with $V^*$ until the point when $V^*$ sends first $V_1$ message for some session $j$. Informally, this is the point when $V^*$ completes the "Knowledge Commitment Protocol" for the first time. So $F_{n,1}$ will keep sending the first message of the sessions requested by $V^*$ and wait for it to respond for one of the sessions. When $V^*$ sends $V_1$ message for some session, $F_{n,1}$ outputs the message of $V^*$. More formally,

**Step 1:** $F_{n,1}$ sets $\gamma = (x, y, R_2)$ and $Msg' = (q_1, g_1, g_1^{a_1})$. $F_{n,1}$ runs $V^*$ on $(Msg', 1, \gamma)$.

**Step 2:** Let output of $V^*$ be $(Msg, j, \gamma, \ell)$. Now it does case analysis on $Msg$.

**Step 2a:** If $Msg = \epsilon$, set $Msg' = (q_\ell, g_\ell, g_\ell^{a_\ell})$ and run $V^*$ on $(Msg', \ell, \gamma)$. Go to Step 2.

**Step 2b:** If $Msg = V_1$ message of session $j$, i.e. $Msg = (B_j, X_j)$, $F_{n,1}$ outputs $(j, B_j, X_j)$.

**Step 2c:** If $Msg = (\text{END}, \text{output})$, $F_{n,1}$ outputs $(\text{END}, \text{output})$.

Note that since $F_{n,1}$ stops whenever $V^*$ sends the $V_1$ message of any session, the only inputs to $V^*$ are prover's $P_1$ message.

Now that we have defined $F_{n,1}$, we define our admissible adversary $\{C_{n,1}, M_{n,1}, B_{n,1}, \text{aux}_n\} = (F_{n,1}, F_{n,1}, \epsilon, \text{aux}_n)$. Now we describe $\{C'_{n,1}, M'_{n,1}, B'_{n,1}, \text{aux}_n\}$ as follows: By $m$-KEA, there must exist an extractor circuit $E_{n,1}$ which takes a subset of the wires of $M_{n,1}$ as input and outputs $(j, b_j)$ such that if $(g_j^{a_j}, B_j, X_j) \in \mathcal{DH}$ then $B_j = g_j^{b_j}$ with all but negligible probability. Here, without loss of generality, for ease of notation, we have assumed that $E_{n,1}$ also outputs $j$ along with $b_j$. This can be done by just using output wires of $M_{n,1}$. Then $\{C'_{n,1}, M'_{n,1}, B'_{n,1}, \text{aux}_n\} = (C_{n,1} \cup E_{n,1}, M_{n,1}, E_{n,1}, \text{aux}_n)$.

We now describe $\{C_{n,i+1}, M_{n,i+1}, B_{n,i+1}, \text{aux}_n\}$ and $\{C'_{n,i+1}, M'_{n,i+1}, B'_{n,i+1}, \text{aux}_n\}$ recursively. Informally, $\{C_{n,i+1}, M_{n,i+1}, B_{n,i+1}, \text{aux}_n\}$ would be a result of polysize malicious extensions to $\{C'_{n,i}, M'_{n,i}, B'_{n,i}, \text{aux}_n\}$ using an extension circuit $F_{n,i+1}$. Here, $F_{n,i+1}$ would continue the simulation using the output of $B'_{n,i}$. It would start by checking if the last benign extraction was successful. If the extraction failed, it outputs SimAbort. Otherwise, it continues simulation till the point when $V^*$ responds with next $V_1$ message for some session $j$. Then $\{C'_{n,i+1}, M'_{n,i+1}, B'_{n,i+1}, \text{aux}_n\}$ would do the benign extraction for session $j$.

**Admissible Adversary:** $\{C_{n,i+1}, M_{n,i+1}, B_{n,i+1}, \text{aux}_n\}$ for some $i \in \{1, 2, \ldots, m\}$.

**Input:** $(x, y, (q_1, g_1, g_1^{a_1}), (q_2, g_2, g_2^{a_2}), \ldots, (q_m, g_m, g_m^{a_m}))$ and $(R_1, R_2)$, where $x \in L$, $y$ is the auxiliary input of length $m$ and $(q_i, g_i) \in L_{QG}$, for all $i$. $R_1$ is the random tape for $C_{n,1}$ and $R_2$ is the random tape for $V^*$.

**Output:** $(j, B_j, X_j)$ or $(\text{END}, \text{output})$ or SimAbort.

**Description:** $\{C_{n,i+1}, M_{n,i+1}, B_{n,i+1}, \text{aux}_n\}$ is the result of polysize malicious extension to $\{C'_{n,i}, M'_{n,i}, B'_{n,i}, \text{aux}_n\}$. Let $F_{n,i+1}$ be this malicious extension. It will simulate the interaction with $V^*$ from the point when $V^*$ sends $i^{th}$ $V_1$ message till $V^*$ sends one more $V_1$ message for some session $j$. These messages would be simulated with the help of the extractions done by the benign part of the circuit $B'_{n,i}$ so far. When $V^*$ sends $V_1$ message for session $j$, then $F_{n,i+1}$ stops and outputs the message of $V^*$. More formally, $F_{n,i+1}$ is defined as follows:

**Step 1:** If $\{C'_{n,i}, M'_{n,i}, B'_{n,i}, \text{aux}_n\}$ outputs SimAbort or $(\text{END}, \text{output})$, then $F_{n,i+1}$ outputs the same. Else find the last output from $V^*$ in $M_{n,i}$. It would be of the form $(j, B_j, X_j, \gamma)$. Set $Msg = (j, B_j, X_j)$ and do the following:

- If $e_j(g_j^{a_j}, B_j) \neq e_j(X_j, g_j)$ then add $(\text{Abort}, j)$ to Aborted. Set $Msg' = (\text{Abort}, j)$ and run $V^*$ on $(Msg', j, \gamma)$. Go to Step 2.
- Find the corresponding output $(j, b_j)$ of $B'_{n,i}$. If not found or if $B_j \neq g_j^{b_j}$, $F_{n,i+1}$ outputs SimAbort.
- If the extraction was successful, $F_{n,i+1}$ knows the discrete log of $B_j$, and it can equivocate in the commitment scheme $\text{Com}_{DL_j}$. Set $Z_j = \text{Com}_{DL_j}(0, \tilde{r}'_j)$. Run $S_{HV}$ on input $x$ to get the view of $\overline{V}$ for session $j$, say $(\text{CMT}_j, \text{CH}_j, \text{RSP}_j)$. Set $Msg' = (\text{CMT}_j, Z_j)$ and run $V^*$ on $(Msg', j, \gamma)$.

**Step 2:** Let output of $V^*$ be $(Msg, j, \gamma, \ell)$ for some $j$ and $\gamma$. Now $F_{n,i+1}$ does case analysis on $Msg$.

**Step 2a:** If $(\text{Abort}, j) \in \text{Aborted}$, Set $Msg' = (\text{Abort}, j)$ and $\text{next} = (Msg', j, \gamma)$.

**Step 2b:** If $Msg = V_1$ message of session $j$, i.e. $Msg = (B_j, X_j)$, then $F_{n,i+1}$ outputs $(j, B_j, X_j)$.

**Step 2c:** If $Msg = V_2$ message of session $j$, i.e. $Msg = \beta_j$, then find $\text{CH}_j$, $\text{RSP}_j$ and $\tilde{r}'_j$ in $M'_{n,i} \cup F_{n,i+1}$ and set $\alpha_j = \text{CH}_j \oplus \beta_j$. Set $\tilde{r}_j = \text{Open}_{DL_j}(0, \alpha_j, \tilde{r}'_j, b_j)$. Set $Msg' = (\alpha_j, \tilde{r}_j, \text{RSP}_j)$ and $\text{next} = (Msg', j, \gamma)$.

**Step 2d:** If $Msg = \epsilon$, then set $Msg' = (q_\ell, g_\ell, g_\ell^{a_\ell})$ and $\text{next} = (Msg', \ell, \gamma)$.

**Step 2e:** If $Msg = (\text{END}, \text{output})$, $F_{n,i+1}$ outputs $(\text{END}, \text{output})$.

**Step 3:** Run $V^*$ on next and go to Step 2.

With the above description of $F_{n,i+1}$ complete, we now define $\{C_{n,i+1}, M_{n,i+1}, B_{n,i+1}, \text{aux}_n\} = (C'_{n,i} \cup F_{n,i+1}, M'_{n,i} \cup F_{n,i+1}, B'_{n,i}, \text{aux}_n)$.

Now that we have defined $\{C_{n,i+1}, M_{n,i+1}, B_{n,i+1}, \text{aux}_n\}$, we define our admissible adversary $\{C'_{n,i+1}, M'_{n,i+1}, B'_{n,i+1}, \text{aux}_n\}$ as follows: By $m$-KEA, there must exist a circuit $E_{n,i+1}$ which takes a subset of internal wires of $M_{n,i+1}$ as input and outputs $(j, b_j)$ such that if $(g_j^{a_j}, B_j, X_j) \in \mathcal{DH}$ then $B_j = g_j^{b_j}$ with all but negligible probability. Here again, without loss of generality, for ease of notation, we assume that $E_{n,i+1}$ also outputs $j$ along with $b_j$. This can be done by just using output wires of $M_{n,i+1}$. Then, we define $\{C'_{n,i+1}, M'_{n,i+1}, B'_{n,i+1}, \text{aux}_n\} = (C_{n,i+1} \cup E_{n,i+1}, M_{n,i+1}, B_{n,i+1} \cup E_{i+1}, \text{aux}_n)$.

Now that we have defined this sequence of admissible adversaries, we will describe our simulator $\mathcal{S}_{V^*}$ in terms of these machines.

**Circuit:** $\mathcal{S}_{V^*}$.
**Input:** $(x, y)$, where $x \in L$ and $y \in \text{aux}_n$ is the auxiliary input of length $m$.
**Output:** View of $V^*$.

**Step 1:** If $V^*$ starts $m$ sessions then $\mathcal{S}_{V^*}$ generates $(q_i, g_i) \xleftarrow{\$} L_{QG}$ for all $i \in \{1, 2, \ldots, m\}$. Each $q_i$ is of length $n$.

**Step 2:** $\mathcal{S}_{V^*}$ generates $a_1, a_2, \ldots, a_m$ uniformly at random such that $a_i \in \mathbb{Z}_{q_i}^*$ and computes $A_i = g_i^{a_i}$ for all $i$.

**Step 3:** $\mathcal{S}_{V^*}$ executes the admissible adversary circuit $(C_{n,m+1}, M_{n,m+1}, B_{n,m+1})$ with the inputs $(x, y, (q_1, g_1, g_1^{a_1}), (q_2, g_2, g_2^{a_2}), \ldots, (q_m, g_m, g_m^{a_m}))$ and $(R_1, R_2)$, where $x \in L$, $y$ is the auxiliary input of $V^*$ of length $m$ and $R_2$ is the random tape of $V^*$ and $R_1$ are the random coins for $C_{n,m+1}$.

**Step 4a:** If $(C_{n,m+1}, M_{n,m+1}, B_{n,m+1})$ outputs SimAbort then $\mathcal{S}_{V^*}$ also outputs SimAbort.

**Step 4b:** If $(C_{n,m+1}, M_{n,m+1}, B_{n,m+1})$ runs to completion with output (END, output), then $\mathcal{S}_{V^*}$ outputs output.

**Theorem B.1.** *If there are $m$ concurrent sessions of $\Pi$ and if our family of admissible adversaries $\mathcal{A}$ contains all polynomial size adversaries and allows polysize malicious extensions, then under $m$-KEA and honest verifier zero-knowledge property of $\langle \overline{P}, \overline{V} \rangle$, the following distribution ensembles are computationally indistinguishable:*

$$\{S_{V^*}(x, y)\}_{m, x \in L, y \in \{0,1\}^m} \text{ and } \{\langle P(x, w), V^*(x, y)\rangle\}_{m, x \in L, w \in W_L(x), y \in \{0,1\}^m}$$

**Proof:** We will prove indistinguishability by a sequence of hybrids. If there are $m$ sessions we will consider $3m + 1$ hybrids, $\mathcal{H}_0 \cup \{\mathcal{H}_{i,1}, \mathcal{H}_{i,2}, \mathcal{H}_{i,3}\}$, for all $i \in [m]$. We will now describe the hybrids in detail. We will assume that all the hybrids also have the witness $w$ for the fact $x \in L$.

- $\mathcal{H}_0$ is the honest hybrid. It runs Step 1 and 2 of $\mathcal{S}_{V^*}$ and then builds $F_{n,1}$ but does not stop on receiving the first response from $V^*$. Instead it uses the witness and interacts honestly in all the sessions. $\mathcal{H}_0 = \{F_{n,1}, F_{n,1}, \epsilon, \text{aux}_n\}$. This hybrid is same as the honest prover interacting with $V^*$.

  In each of the following hybrids we build on the admissible adversary circuit $\{C'_{n,i}, M'_{n,i}, B'_{n,i}, \text{aux}_n\}$. If its malicious part $M'_{n,i}$ outputs a tuple, we will call its session number $j$.

- $\mathcal{H}_{i,1}$ runs Step 1 and 2 of $\mathcal{S}_{V^*}$ and then builds $\{C'_{n,i}, M'_{n,i}, B'_{n,i}, \text{aux}_n\}$ with the inputs $(x, y, (q_1, g_1, g_1^{a_1}), (q_2, g_2, g_2^{a_2}), \ldots, (q_m, g_m, g_m^{a_m}))$ and $(R_1, R_2)$, where $x \in L$, $y$ is the auxiliary input of $V^*$ of length $m$ and $R_2$ is the random tape of $V^*$ and $R_1$ are the random coins of $\{C'_{n,i}, M'_{n,i}, B'_{n,i}, \text{aux}_n\}$. If $\{C'_{n,i}, M'_{n,i}, B'_{n,i}, \text{aux}_n\}$ outputs SimAbort, then $\mathcal{H}_{i,1}$ does the same. Else find the last output from $V^*$ in $M_{n,i}$. It would be of the form $(j, B_j, X_j, \gamma)$. Now, start building polysize malicious extension $F_{n,i+1}$ to $\{C'_{n,i}, M'_{n,i}, B'_{n,i}, \text{aux}_n\}$. $F_{n,i+1}$ does the following tests:

  – If $e_j(g_j^{a_j}, B_j) \neq e_j(X_j, g_j)$ then add (Abort, $j$) to Aborted.
  – Find the corresponding output $(j, b_j)$ of $B'_{n,i}$. If not found or $B_j \neq g_j^{b_j}$, $F_{n,i+1}$ outputs SimAbort.

If the tests pass, $F_{n,i+1}$ continues as follows: Among the unaborted sessions, for all the sessions $\ell$ such that $\ell \neq j$ and $(\ell, b_\ell)$ lies in the output of $B'_{n,i}$, it uses the extracted values to simulate the sessions. For rest of the sessions, it uses the witness to generate the messages honestly. Note that though $\mathcal{H}_{i,1}$ has the extracted value for session $j$, it does not use it and acts honestly in that session. $\mathcal{H}_{i,1} = (C'_{n,i} \cup F_{n,i+1}, M'_{n,i} \cup F_{n,i+1}, B'_{n,i}, \text{aux}_n)$.

- $\mathcal{H}_{i,2}$ is same as $\mathcal{H}_{i,1}$ with the following change. It chooses $\alpha_j \xleftarrow{\$} \mathbb{Z}_q$, but sets $Z_j = \text{Com}_{DL_j}(0; \tilde{r}'_j)$. Later while opening it sets $\tilde{r}_j = \text{Open}_{DL_j}(0, \alpha_j, \tilde{r}'_j, b_j)$ and opens the commitment to $\alpha_j$ and $\tilde{r}_j$. It generates all other messages of session $j$ honestly.

  The hybrids $\mathcal{H}_{i,1}$ and $\mathcal{H}_{i,2}$ are identical because the commitment scheme $\text{Com}_{DL_j}$ is perfectly hiding and hence, $\text{Com}_{DL_j}(0)$ and $\text{Com}_{DL_j}(\alpha_j)$ are identically distributed.

- $\mathcal{H}_{i,3}$ does the following change in $\mathcal{H}_{i,2}$. While generating $P_2$ message of session $j$, it runs $S_{HV}$ to get $(\text{CMT}_j, \text{CH}_j, \text{RSP}_j)$. It sends $(\text{CMT}_j, \text{Com}_{DL_j}(0; \tilde{r}'_j))$ as $P_2$ message. For $P_3$ message, it sets $\alpha_j = \beta_j \oplus \text{CH}_j$, $\tilde{r}_j = \text{Open}_{DL_j}(0, \alpha_j, \tilde{r}'_j, b_j)$ and sends $(\alpha_j, \tilde{r}_j, \text{RSP}_j)$. Hence, in $\mathcal{H}_{i,3}$ all sessions $\ell$ such that $(\ell, b_\ell) \in$ output of $B^j_{n,i}$ are simulated and rest all sessions are honest. Note that $\mathcal{H}_{m,3}$ is same as the interaction between $\mathcal{S}_{V^*}$ and $V^*$.

  Below we will prove the indistinguishability of $\mathcal{H}_{i,2}$ and $\mathcal{H}_{i,3}$.

  First note that the only difference between the hybrids $\mathcal{H}_{i,3}$ and $\mathcal{H}_{i+1,1}$ is that in $\mathcal{H}_{i+1,1}$ we add $E_{n,i+1}$ to $B'_{n,i}$. But SimAbort happens in $\mathcal{H}_{i+1,1}$ and not in $\mathcal{H}_{i,3}$ only when $V^*$ sends correctly formed $(B, X)$ but we fail to extract the correct value in $E_{n,i+1}$. This probability is negligible by $m$-KEA. Hence, the hybrids $\mathcal{H}_{i,3}$ and $\mathcal{H}_{i+1,1}$ are statistically close.

In order to show indistinguishability between $\mathcal{H}_0$ and $\mathcal{H}_{m,3}$, we are just left with showing indistinguishability between $\mathcal{H}_{i,2}$ and $\mathcal{H}_{i,3}$. We will show this by contradiction. Let us assume there is a distinguisher $\mathcal{D}$ which distinguishes between $\mathcal{H}_{i,2}$ and $\mathcal{H}_{i,3}$ for some $i$ and auxiliary input $y$. Then we will show a distinguisher $\mathcal{D}'$ for

$$\{S_{HV}(x)\}_{x \in L} \text{ and } \{\langle \overline{P}(x,w), \overline{V}(x)\rangle\}_{x \in L, w \in W_L(x)}.$$

This would contradict the honest verifier zero-knowledge property of $\langle \overline{P}, \overline{V} \rangle$. $\mathcal{D}'$ is given a 3-round transcript $\mathcal{T}' = \{\text{CMT}, \text{CH}, \text{RSP}\}$ as input for the NP-statement $x \in L$ which is either for $\overline{P}$ or $S_{HV}$ for the canonical argument system. Using the witness $w$ for $x \in L$, $\mathcal{D}'$ generates an input $\mathcal{H}_I$ for $\mathcal{D}$ which is same as $\mathcal{H}_{i,3}$ except for the following change. Instead of running $S_{HV}$ for the session $j$ (defined above), it uses $\{\text{CMT}, \text{CH}, \text{RSP}\}$ for that session.

If the input to $\mathcal{D}'$ is from $\{\langle \overline{P}(x,w), \overline{V}(x)\rangle\}_{x \in L, w \in W_L(x)}$, the input to $\mathcal{D}$ is identical to $\mathcal{H}_{i,2}$. This is because since honest verifier's CH is distributed uniformly in $\{0,1\}^n$, $\alpha_j$ in $\mathcal{H}_I$ will be distributed uniformly. On the other hand, if the input of $\mathcal{D}'$ is from $\{S_{HV}(x)\}_{x \in L}$, the input to $\mathcal{D}$ is identical to $\mathcal{H}_{i,3}$. So, if $\mathcal{D}$ says that $\mathcal{H}_I$ is distributed identically to $\mathcal{H}_{i,2}$, then $\mathcal{D}'$ says that $\mathcal{T}'$ is generated by $\overline{P}$. Else, it is generated by $S_{HV}$.

The success probability of $\mathcal{D}'$ in distinguishing between transcripts of $\overline{P}$ and $S_{HV}$ is same as the success probability of $\mathcal{D}$ in distinguishing $\mathcal{H}_{i,2}$ and $\mathcal{H}_{i,3}$. Hence, $\mathcal{D}'$ distinguishes between the following two ensembles with non-negligible probability

$$\{S_{HV}(x)\}_{x \in L} \text{ and } \{\langle \overline{P}(x,w), \overline{V}(x)\rangle\}_{x \in L, w \in W_L(x)}$$

in contradiction to property **B3** of the canonical argument system.

## C  Size of the Simulator Circuit

**Theorem C.1.** *Under m-KEA, there exists a constant $c$ such that the size of the circuit of $\mathcal{S}_{V^*}$, denoted by $|\mathcal{S}_{V^*}|$, is bounded by $n^c \cdot |V^*|^c + \text{poly}(n)$.*

**Proof:** Let $m$-KEA hold w.r.t. a family of admissible adversaries $\mathcal{A}$. Then in Section 6, the simulator is described by a circuit $S = \{C_{n,m+1}, M_{n,m+1}, B_{n,m+1}, \text{aux}_n\}$.

Now, $|\mathcal{S}_{V^*}| = |C_{n,m+1}| + |C|$, where $C$ is the circuit which generates $(p_i, q_i, g_i)$ at random from $L_{PQG}$, generates $a_i$ at random from $\mathbb{Z}_q^*$ and computes $g_i^{a_i}$ for all $i \in [m]$. There exists a fixed constant $c_1$ such that $|C| \leq n^{c_1}$. We are left with computing $|C_{n,m+1}|$.

Since, $M_{n,m+1}$ and $B_{n,m+1}$ form a partition of $C_{n,m+1}$, $|C_{n,m+1}| = |M_{n,m+1}| + |B_{n,m+1}|$. By construction, $B_{n,m+1} = \bigcup_{i=1}^{m} E_{n,i}$. By $m$-KEA, there exists a constant $c' > 0$, such that $|E_{n,i}| \leq (n \cdot |M_{n,i}|)^{c'}$. So,

$$|B_{n,m+1}| = \sum_i^m |E_{n,i}| \leq \sum_i^m (n \cdot |M_{n,i}|)^{c'} \leq n^{c'} \cdot \sum_i^m |M_{n,m}|^{c'} \leq n^{c'} \cdot m \cdot |M_{n,m}|^{c'}.$$

Using the above, we get $|C_{n,m+1}| \leq |M_{n,m+1}| + n^{c'} \cdot m \cdot |M_{n,m}|^{c'} \leq n^{c_2} \cdot |M_{n,m+1}|^{c_2}$ for some constant $c_2 > c'$.

$M_{n,m+1}$ calls $V^*$ at most $3 \cdot m$ times, calls $S_{HV}$ at most $m$ times and generates all other messages using a circuit of size at most $n^{c_3}$ for some constant $c_3$. Hence, $|M_{n,m+1}| \leq 3m \cdot |V^*| + m \cdot |S_{HV}| + n^{c_3}$. We also know that if $n$ is the security parameter then $|S_{HV}| \leq n^{c_4}$ for some constant $c_4 > 0$. We get $|M_{n,m+1}| \leq 3 \cdot m \cdot |V^*| + n^{c_5}$. Combining all we get,

$$|\mathcal{S}_{V^*}| = |C_{n,m+1}| + |C| \leq n^{c_2} \cdot |M_{n,m+1}|^{c_2} + n^{c_1} \leq n^{c_2} \cdot (3 \cdot m |V^*| + n^{c_5})^{c_2} + n^{c_1} \leq n^c \cdot |V^*|^c + poly(n),$$

where $c > 0$ is a fixed constant.

# D  Constant Round Protocol for Concurrent Zero-Knowledge using Knowledge of Exponent Assumption in General Groups

The constant round protocol described below is a concurrent zero-knowledge protocol under knowledge of exponent assumption $m$-KEA in general groups. The assumption is similar to that described in Section 3.4, but now it is assumed to hold w.r.t. general groups of prime order (see Assumption D.5). This protocol is very similar to the previous protocol apart from the following change: In the previous protocol, when the verifier replied back with $(B, X)$ on input $(g, A)$, the prover checked if $B^a = X$. Here $a$ is the discrete log of $A$, which is known to the prover. In the protocol described in this section, the prover will not do any such check. Instead, the verifier will prove in zero-knowledge that indeed there exists a '$b$' such that $B = g^b$ and $X = A^b$ using a constant round statistically sound zero-knowledge protocol $\Pi_{ZK}$. Such a protocol was given by Goldreich and Kahan [GK96] for all of NP, but more efficient such protocols exist for proving Diffie-Hellman pairs and can be used (see e.g. [Gol01] and the references therein).

We start by giving a few additional definitions.

**Definition D.1.** Let $L_{PQG}$ denote the set $\{(p, q, g)\}$ of primes and generators, where $p$ and $q$ are primes such that $p = 2q + 1$ and $g$ is an element of order $q$ in $Z_p^*$.

**Definition D.2.** Let $\Pi_{ZK}$ be a constant round statistically sound zero-knowledge protocol for all NP. We will use the protocol given by Goldreich and Kahan [GK96].

**Knowledge Assumption:** Below, by a circuit $C$ we mean a collection of Boolean gates and wires. We use the non-standard convention that certain gates are specially marked as output gates.

**Definition D.3** (Admissible family of Adversaries)**.** An admissible family of adversaries $\mathcal{A}$ is a family of sets such that the following properties hold: Each set $S \in \mathcal{A}$ is such that $S = \{C_n, M_n, B_n, \mathsf{aux}_n\}_{n \in \mathbb{N}}$. For each such set $S$, there exist constants $c, c' > 0$, such that $C_n$ is a circuit with $|C_n| \leq n^c$, and $\mathsf{aux} \subseteq \{0,1\}^{n^{c'}}$. Furthermore, $\{M_n, B_n\}$ is a partition of the gates and the wires of the circuit $C_n$. If $x$ is the input to $C_n$ then by $M_n(x)$ we refer to the result of the computation $C_n(x)$ restricted to the output wires in $M_n$; we define $B_n(x)$ similarly.

We will refer to $M_n$ and $B_n$ as the malicious and the benign parts respectively of the adversary circuit $C_n$.

**Definition D.4** (*$\mathcal{A}$ admits polysize malicious extensions*)**.** An admissible family of adversaries $\mathcal{A}$ *admits polysize malicious extensions* if the following holds: For any set of circuits $S \in \mathcal{A}$ where $S = \{C_n, M_n, B_n, \mathsf{aux}_n\}_{n \in \mathbb{N}}$, and any polysize circuit family $\{F_n\}_{n \in \mathbb{N}}$ such that $\exists d > 0$, $|F_n| < n^d$ and the input wires to $F_n$ are a subset of the wires in $M_n$ (including both internal and output wires) and the output wires of $B_n$, we have that $S' = \{C_n \cup F_n, M_n \cup F_n, B_n, \mathsf{aux}_n\} \in \mathcal{A}$.

Next, based on the definition above, we define a variant of knowledge of exponent assumption based on the one described by Hada and Tanaka [HT98].

**Assumption D.5.** [$m$-Knowledge of Exponent Assumption ($m$-KEA) w.r.t. admissible adversaries] We say that the $m$-Knowledge of Exponent Assumption holds with respect to a family of admissible adversaries $\mathcal{A}$, if for every $c > 0$, there exists a constant $c' > 0$ such that the following holds: For $m = n^c$, fix any $S = \{C_n, M_n, B_n, \mathsf{aux}_n\}_{n \in \mathbb{N}} \in \mathcal{A}$. Then there exists a family of extraction circuits $\{E_n\}_{n \in \mathbb{N}}$ whose inputs are a subset of any wires in $M_n$, such that $|E_n| \leq (n \cdot |M_n|)^{c'}$. (Informally, this condition requires that the extraction only uses the internal wires of the malicious part of the adversary.) Furthermore, we require that the following conditions hold:

1. For all sufficiently large $n$, every polynomial $poly(\cdot)$, the following is true for all $aux \in \mathsf{aux}_n$: Consider the following probabilistic experiment: For $i \in [1, m]$, primes $p_i, q_i$ and generators $g_i$ are chosen randomly such that $(p_i, q_i, g_i) \in L_{PQG}$, where $p_i$ is chosen to be of length $n$. Values $a_1, \ldots, a_m$ are chosen at random such that $a_i \in \mathbb{Z}^*_{q_i}$. Finally, $R$ is chosen uniformly at random from sufficiently long strings so that the length of the tuple $x = ((p_1, q_1, g_1, g_1^{a_1}), \ldots, (p_m, q_m, g_m, g_m^{a_m}), aux, R)$ is exactly the length of the input to circuit $C_n$. If the input to $C_n$ is not long enough to allow such an input then the assumption is vacuously true for this $S$. Now, we consider the output of $M_n(x)$, which we interpret as a tuple $(j, B, X)$, where $j \in [m]$, and both $B$ and $X$ are in $\mathbb{Z}_p$. Then, we interpret the output of $E_n(x)$ as the value $b_j \in \mathbb{Z}_q$, and require the following to be true:

$$Pr\left[X = B^{a_j} \wedge B \neq g_j^{b_j}\right] < \frac{1}{poly(n)}.$$

    (Informally, this condition states that if the malicious part of the adversary outputs a tuple so that $(g_j, g_j^{a_j}, B, X)$ form a Diffie-Hellman tuple, then the extractor $E_n$ successfully outputs the discrete log of $B$ with respect to $g_j$.)

2. We have that $(C_n \cup E_n, M_n, B_n \cup E_n, \mathsf{aux}_n) \in \mathcal{A}$. (Informally, this means that the extraction circuit created by this assumption is benign.)

**Definition D.6.** An admissible set of adversaries $\mathcal{A}$ *contains all polysize malicious adversaries* if for all $c, c' > 0$, and for all circuit families $\{C_n\}_{n \in \mathbb{N}}$ such that $|C_n| \leq n^c$, for each $n$ there exists some subset $\mathsf{aux}_n \subseteq \{0,1\}^{n^{c'}}$, such that $(C_n, C_n, \epsilon, \mathsf{aux}_n) \in \mathcal{A}$. We say that $\mathcal{A}$ *contains all polysize malicious adversaries with all polysize auxiliary inputs* if $\mathsf{aux}_n = \{0,1\}^{n^{c'}}$ for each circuit family above.

**Theorem D.7** (Informal)**.** *If the m-Knowledge of Exponent assumption holds with respect to an admissible adversary family $\mathcal{A}$ such that $\mathcal{A}$ contains all polysize malicious circuits and allow polysize malicious extension, and DHLA holds, then there exist constant-round concurrent zero-knowledge arguments for **NP** in the plain model.*

*Furthermore, if $\mathcal{A}$ contains all polysize malicious adversaries with all polysize auxiliary inputs, then there exist constant-round concurrent zero-knowledge arguments for **NP** in the plain model with respect to arbitrary auxiliary inputs.*

## D.1 Protocol Description

The protocol starts by asking the verifier to use Knowledge Commitment Protocol to commit to a value $b$ in $B = g^b$. Then the verifier proves that this commitment is correctly generated using $\Pi_{ZK}$. Following this, we use equivocal commitments whose trapdoor is $b$ to run a coin flipping protocol between the prover and the verifier. In parallel with the coin flipping protocol, we run a parallel repetition of Blum's Hamiltonicity protocol, where the result of coin flipping protocol determines the challenge message. We describe the constant-Round protocol for concurrent zero-knowledge argument in Figure 3.

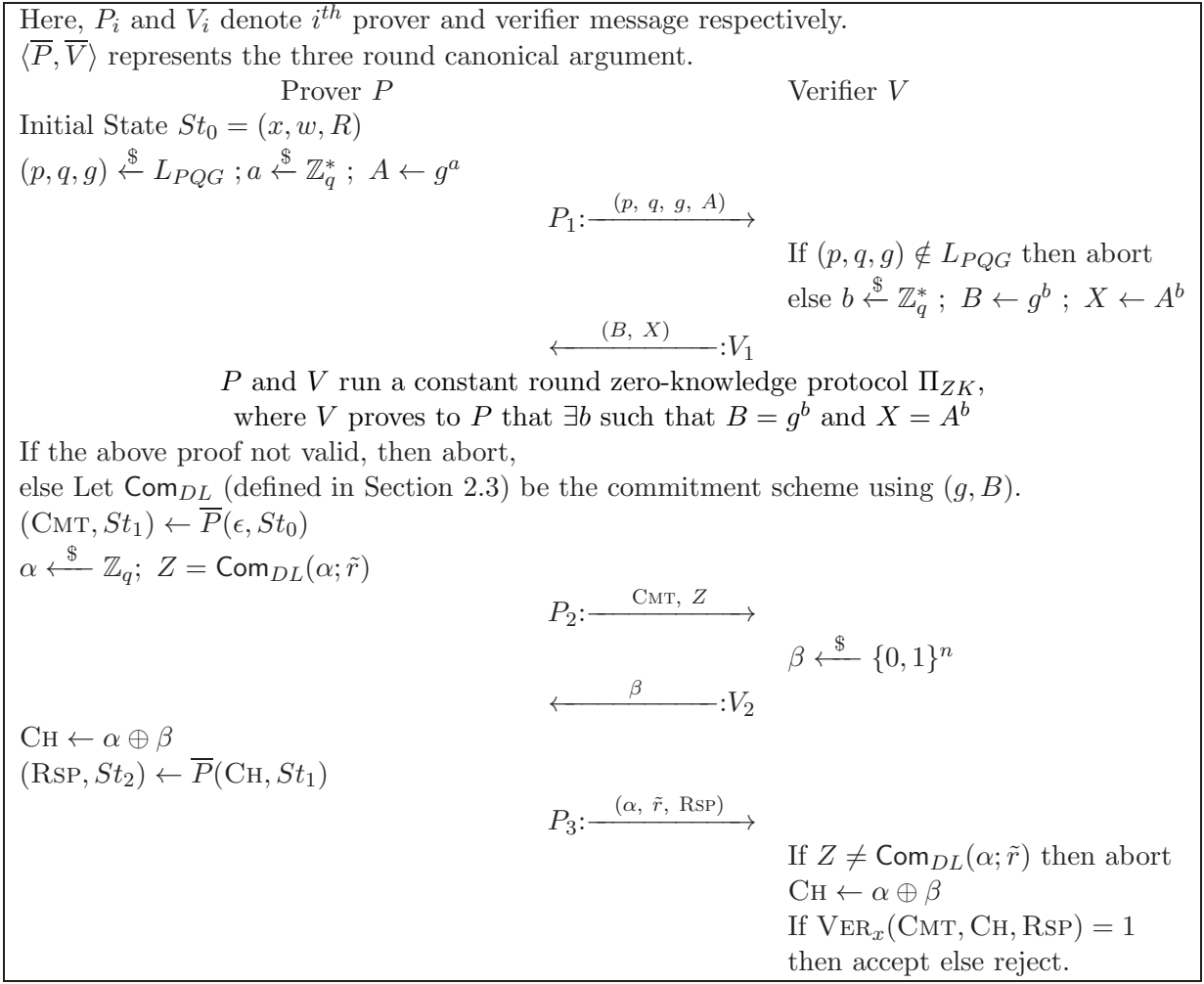

Figure 3: $\Pi$: Constant Round Protocol for $c\mathcal{ZK}$ $(P,V)$

This protocol uses the discrete log based commitment scheme $\mathsf{Com}_{DL}$ which is binding under the hardness of *DHLA*. The secret value $b$ committed to by the verifier satisfies the following properties.

**R1:** For **Soundness:** Under *DHLA* (Assumption 3.1) and zero-knowledge property of $\Pi_{ZK}$, any cheating prover while interacting with the honest verifier cannot get the secret coins of the verifier. Hence, any cheating prover cannot output the discrete log of $B$ sent by the verifier in Figure 2.

**R2:** For **Zero-knowledge:** Under *m-KEA* (Assumption D.5), our simulator will be able to output the discrete log of $B$ no matter how the verifier behaves. Once the simulator gets the secret coins of $V^*$, which is the trapdoor to equivocal commitment scheme, the simulation is easy.

For **R2**, informally, it seems that even the cheating verifier must start by simply choosing $b$ and computing $(g^b, A^b)$ in order to pass the check $X = B^a$. That is, we assume that verifier knows the secret coins $b$ whenever it manages to convince the prover in $\Pi_{ZK}$. $m$-KEA defined in Assumption D.5 captures this idea of knowledge and knowledge extraction formally. Under this variant of knowledge of exponent assumption, we will design a simulator which will extract the secret coins of the cheating verifier. Since, the simulator will have the trapdoor to $\mathsf{Com}_{DL}$, it will be able to equivocate on its commitment to $\alpha$ just as before and force the outcome of the coin flipping protocol to the challenge string output by the honest verifier simulator $S_{HV}$.

## D.2  $\Pi$ is Computationally Sound

We prove soundness of $\Pi$ by the two steps used in proving the soundness of the previous protocol. Let $S_{ZK}$ be the black-box zero-knowledge simulator for $\Pi_{ZK}$ and $P^*$ denote the cheating prover.

- If $P^*$ succeeds in equivocating its commitment in coin flipping protocol then we can extract the trapdoor value $b$ of Knowledge Commitment Protocol from $P^*$. This shows that $P^*$ can be used to efficiently compute $b$ and thereby break *DHLA*.

- We show that if $P^*$ does not equivocate on its commitment in coin flipping protocol and convinces the verifier of a false statement, then such a $P^*$ can be used to violate the underlying strong soundness of canonical arguments. In other words, it would violate the underlying soundness of Blum's Hamiltonicity protocol.

We prove the first step by a sequence of two lemmas. Let $\mathcal{G}$ be the following interactive game similar to that defined in Section 5.

1. $Sim$ runs the above protocol with $P^*$ (including honest execution of $\Pi_{ZK}$) till $P^*$ commits to $\alpha$ using random coins $\tilde{r}$ in the above protocol using commitment scheme $\mathsf{Com}_{DL}$ as defined before. $P^*$ sets $Z = \mathsf{Com}_{DL}(\alpha; \tilde{r})$ and sends $Z$ to $Sim$.

2. $Sim$ sends $\beta$ to $P^*$.

3. $P^*$ sends $(\alpha_1, \tilde{r}_1, \textsc{Rsp})$ to $Sim$ such that $Z = \mathsf{Com}_{DL}(\alpha_1; \tilde{r}_1)$.

4. $Sim$ rewinds $P^*$ to Step 2 and sends it $\beta'$ such that $\beta' \neq \beta$. $P^*$ wins if it sends $(\alpha_2, \tilde{r}_2, \textsc{Rsp})$ to $Sim$ such that $Z = \mathsf{Com}_{DL}(\alpha_2; \tilde{r}_2)$ and $\alpha_1 \neq \alpha_2$.

Let $\mathcal{G}'$ be a modified game in which $Sim$ runs $S_{ZK}$ to simulate the proof in first step instead of using the witness.

**Lemma D.8.** $Pr[P^* \text{ wins } \mathcal{G}] - Pr[P^* \text{ wins } \mathcal{G}']$ is negligible.

**Proof:** We will prove this by contradiction. If there is a non-negligible function $\gamma(n)$ such that $Pr[P^* \text{ wins } \mathcal{G}] - Pr[P^* \text{ wins } \mathcal{G}'] > \gamma(n)$, then we can construct a distinguisher $D$ which breaks the ZK-property of $\Pi_{ZK}$ as follows: Let $\mathcal{Ch}'$ be the challenger for ZK. $D$ starts the game with $P^*$ and forwards the messages between $P^*$ and $\mathcal{Ch}'$ until the end of $\Pi_{ZK}$. Now $D$ completes the remaining game with $P^*$. If $P^*$ wins, $D$ claims that $\Pi_{ZK}$ was given with actual witness, otherwise $D$ says that $\Pi_{ZK}$ was simulated. It can be shown that success probability of $D$ is $1/2 + \gamma(n)/2$, which is non-negligible. This is a contradiction since $\Pi_{ZK}$ is a zero-knowledge protocol.

**Lemma D.9.** Using above, we prove that no cheating prover can win $\mathcal{G}$ with non-negligible probability. Under *DHLA*, for every probabilistic polynomial time machine $P^*$, every polynomial $poly(\cdot)$, and all sufficiently large $n$'s,

$$Pr[P^* \text{ wins } \mathcal{G}] < \frac{1}{poly(n)}$$

where probability is over choice of $\alpha, \beta$ and coins of $P^*$ and $n$ is the security parameter.

**Proof:** We will prove this by contradiction. If there is a polynomial $f(n)$ such that $Pr[P^* wins \ \mathcal{G}] > 1/f(n)$, then we can construct an adversary $\mathcal{A}$ for *DHLA*. $\mathcal{A}$ runs $P^*$ and gets $(q, g, A)$ and sends $(g, A)$ to the challenger $\mathcal{Ch}$ of *DHLA*. $\mathcal{Ch}$ prepares the challenge tuple by choosing a random $b$ and sends $(B = g^b, X = A^b)$ to $\mathcal{A}$ which it forwards to $P^*$. $\mathcal{A}$ runs $S_{ZK}$ on $P^*$ to simulate $\Pi_{ZK}$. By lemma D.8, the success probability of $P^*$ in winning $\mathcal{G}$ can not decrease non-negligibly when given a simulated proof.

$P^*$ and $\mathcal{A}$ continue running the protocol $\Pi$ until the opening of $Z$ as $\alpha$. After this opening, $\mathcal{A}$ rewinds $P^*$ until the commitment $Z$ and runs $P^*$ again with a different $\beta'$ and looks at the opening of $Z$ by $P^*$. If $P^*$ opens $Z$ to the same $\alpha$, $\mathcal{A}$ aborts. Else if $P^*$ opens $Z$ to a $\alpha'$ such that $\alpha \neq \alpha'$, $\mathcal{A}$ can compute $b$, the discrete log of $B$, as described in Section 2.3. $\mathcal{A}$ sends $b$ to $\mathcal{C}h$.

$Pr[\mathcal{A}\ breaks\ DHLA] = Pr[P^*\ wins\ \mathcal{G}] > 1/f(n) - \epsilon$, where $\epsilon$ is the negligible change in success probability of $P^*$ when $\Pi_{ZK}$ is simulated. This contradicts $DHLA$.

**Theorem D.10.** *Under Lemma D.9 and strong soundness property (B2) of $\langle \overline{P}, \overline{V} \rangle$, protocol $\Pi$ is computationally sound.*

**Proof:** The proof is same as that of Theorem 5.2.

## D.3 Description of the simulator

To describe the simulator for protocol in Figure 3, we again describe a sequence of adversaries. These adversaries are very similar to the ones described before. There is a change in $\{C_{n,i+1}, M_{n,i+1}, B_{n,i+1}, \mathsf{aux}_n\}$ but for the sake of completion, we give the description of all the adversaries.

In the concurrent setting, the verifier may start an unbounded number of sessions with the prover and may interleave them in any way he wants. One such individual session has constant number of rounds (as shown in Figure 3). In this section, we will model our cheating verifier $V^*$ as a next message function with a state $\gamma$.

$$V^*(Msg',\ k,\ \gamma') \to (Msg,\ j,\ \gamma,\ \ell)$$

where $Msg'$ is the prover's (or simulator's) message from the session $k$ and $\gamma'$ is the last state of $V^*$. In response, $V^*$ sends message $Msg$ corresponding to some session $j$ and changes its state to $\gamma$. Prover (or simulator's) next message would be the next message from the session $j$. In case $Msg$ is $\epsilon$, then the verifier is requesting for the first message of session $\ell$. Verifier can also output a special message $(\text{END}, \mathsf{output})$, which means that $V^*$ wants to stop the execution with output $\mathsf{output}$.

To describe our simulator $\mathcal{S}_{V^*}$, we will first describe a sequence of admissible adversaries $\{C_{n,i}, M_{n,i}, B_{n,i}, \mathsf{aux}_n\}$ and $\{C'_{n,i}, M'_{n,i}, B'_{n,i}, \mathsf{aux}_n\}$ for all $i \in \{1, 2, \ldots, m+1\}$. First, we will describe these for $i = 1$ followed by $i > 1$ recursively using $\{C_{n,i-1}, M_{n,i-1}, B_{n,i-1}, \mathsf{aux}_n\}$ and $\{C'_{n,i-1}, M'_{n,i-1}, B'_{n,i-1}, \mathsf{aux}_n\}$. Each of these circuits will maintain and update the set of aborted sessions called Aborted. We will assume that the simulator knows the upper bound on $m$, the number of sessions that $V^*$ executes. Also, whenever $V^*$ stops, our simulator stops with the output of $V^*$.

**Admissible adversary:** $\{C_{n,1}, M_{n,1}, B_{n,1}, \mathsf{aux}_n\}$.

**Input:** $(x, y, (p_1, q_1, g_1, g_1^{a_1}), \ldots, (p_m, q_m, g_m, g_m^{a_m}))$ and $(R_1, R_2)$, where $x \in L$, $y$ is the auxiliary input of length $m$ and $(p_i, q_i, g_i) \in L_{PQG}$, for all $i$. Each $p_i$ is of length $n$.

**Output:** $(j, B_j, X_j)$ or $(\text{END}, \mathsf{output})$.

**Description:** We will start building the circuit $F_{n,1}$ as follows: $F_{n,1}$ will simulate the interaction with $V^*$ until the point when $V^*$ sends first $V_1$ message for some session $j$. Informally, this is the point when $V^*$ completes the "Knowledge Commitment Protocol" for the first time. So $F_{n,1}$ will keep sending the first message of the sessions requested by $V^*$ and wait for it to respond for one of the sessions. When $V^*$ sends $V_1$ message for some session, $F_{n,1}$ outputs the message of $V^*$. More formally,

**Step 1:** $F_{n,1}$ sets $\gamma = (x, y, R_2)$ and $Msg' = (p_1, q_1, g_1, g_1^{a_1})$. $F_{n,1}$ runs $V^*$ on $(Msg', 1, \gamma)$.

**Step 2:** Let output of $V^*$ be $(Msg, j, \gamma, \ell)$. Now it does case analysis on $Msg$.

**Step 2a:** If $Msg = \epsilon$, set $Msg' = (p_\ell, q_\ell, g_\ell, g_\ell^{a_\ell})$ and run $V^*$ on $(Msg', \ell, \gamma)$. Go to Step 2.

**Step 2b:** If $Msg = V_1$ message of session $j$, i.e. $Msg = (B_j, X_j)$, $F_{n,1}$ outputs $(j, B_j, X_j)$.

**Step 2c:** If $Msg = (\text{END}, \mathsf{output})$, $F_{n,1}$ outputs $(\text{END}, \mathsf{output})$.

Note that since $F_{n,1}$ stops whenever $V^*$ sends the $V_1$ message of any session, the only inputs to $V^*$ are prover's $P_1$ message.

Now that we have defined $F_{n,1}$, we define our admissible adversary $\{C_{n,1}, M_{n,1}, B_{n,1}, \text{aux}_n\} = (F_{n,1}, F_{n,1}, \epsilon, \text{aux}_n)$. Now we describe $\{C'_{n,1}, M'_{n,1}, B'_{n,1}, \text{aux}_n\}$ as follows: By $m$-KEA, there must exist an extractor circuit $E_{n,1}$ which takes a subset of the wires of $M_{n,1}$ as input and outputs $(j, b_j)$ such that if $X_j = B_j^{a_j}$ then $B_j = g_j^{b_j}$ with all but negligible probability. Here, without loss of generality, for ease of notation, we have assumed that $E_{n,1}$ also outputs $j$ along with $b_j$. This can be done by just using output wires of $M_{n,1}$. Then $\{C'_{n,1}, M'_{n,1}, B'_{n,1}, \text{aux}_n\} = (C_{n,1} \cup E_{n,1}, M_{n,1}, E_{n,1}, \text{aux}_n)$.

We now describe $\{C_{n,i+1}, M_{n,i+1}, B_{n,i+1}, \text{aux}_n\}$ and $\{C'_{n,i+1}, M'_{n,i+1}, B'_{n,i+1}, \text{aux}_n\}$ recursively. Informally, $\{C_{n,i+1}, M_{n,i+1}, B_{n,i+1}, \text{aux}_n\}$ would be a result of polysize malicious extensions to $\{C'_{n,i}, M'_{n,i}, B'_{n,i}, \text{aux}_n\}$ using an extension circuit $F_{n,i+1}$. Here, $F_{n,i+1}$ would continue the simulation using the output of $B'_{n,i}$. Whenever $\Pi_{ZK}$ phase of any session completes successfully, it checks if the extraction was successful. If the extraction failed, it outputs SimAbort. Otherwise, it continues simulation till the point when $V^*$ responds with next $V_1$ message for some session $j$. Then $\{C'_{n,i+1}, M'_{n,i+1}, B'_{n,i+1}, \text{aux}_n\}$ would do the benign extraction for session $j$.

**Admissible Adversary:** $\{C_{n,i+1}, M_{n,i+1}, B_{n,i+1}, \text{aux}_n\}$ for some $i \in \{1, 2, \ldots, m\}$.
**Input:** $(x, y, (p_1, q_1, g_1, g_1^{a_1}), (p_2, q_2, g_2, g_2^{a_2}), \ldots, (p_m, q_m, g_m, g_m^{a_m}))$ and $(R_1, R_2)$, where $x \in L$, $y$ is the auxiliary input of length $m$ and $(p_i, q_i, g_i) \in L_{PQG}$, for all $i$. Each $p_i$ is of length $n$.
**Output:** $(j, B_j, X_j)$ or (END, output) or SimAbort.
**Description:** $\{C_{n,i+1}, M_{n,i+1}, B_{n,i+1}, \text{aux}_n\}$ is the result of polysize malicious extension to $\{C'_{n,i}, M'_{n,i}, B'_{n,i}, \text{aux}_n\}$. Let $F_{n,i+1}$ be this malicious extension. It will simulate the interaction with $V^*$ from the point when $V^*$ sends $i^{th}$ $V_1$ message till $V^*$ sends one more $V_1$ message for some session $j$. These messages would be simulated with the help of the extractions done by the benign part of the circuit $B'_{n,i}$ so far. When $V^*$ sends $V_1$ message for session $j$, then $F_{n,i+1}$ stops and outputs the message of $V^*$. More formally, $F_{n,i+1}$ is defined as follows:

**Step 1:** If $\{C'_{n,i}, M'_{n,i}, B'_{n,i}, \text{aux}_n\}$ outputs SimAbort or (END, output), then $F_{n,i+1}$ outputs the same. Else find the last output from $V^*$ in $M_{n,i}$. It would be of the form $(j, B_j, X_j, \gamma)$. $B'_{n,i}$ would have attempted to extract the discrete log of $B_j$. Since the next message in Session $j$ is $V^*$'s first message for $\Pi_{ZK}$, run $V^*$ on $(\epsilon, j, \gamma)$.

**Step 2:** Let the output of $V^*$ be $(Msg, j, \gamma, \ell)$ for some $j$ and $\gamma$. Now $F_{n,i+1}$ does case analysis on $Msg$.

**Step 2a:** If $(\text{Abort}, j) \in \text{Aborted}$, Set $Msg' = (\text{Abort}, j)$ and $\text{next} = (Msg', j, \gamma)$.

**Step 2b:** If $Msg = V_1$ message of session $j$, i.e. $Msg = (B_j, X_j)$, then $F_{n,i+1}$ outputs $(j, B_j, X_j)$.

**Step 2c:** If $Msg$ is the last message of $\Pi_{ZK}$ for session $j$, do the following:
- If the proof fails, add $(\text{Abort}, j)$ to Aborted. Set $Msg' = (\text{Abort}, j)$ and $\text{next} = (Msg', j, \gamma)$.
- Find the corresponding output $(j, b_j)$ of $B'_{n,i}$. If not found or if $B_j \neq g_j^{b_j}$, $F_{n,i+1}$ outputs SimAbort.
- If the proof is accepted and there is a valid $(j, b_j)$, $F_{n,i+1}$ knows the discrete log of $B_j$, and it can equivocate in the commitment scheme $\text{Com}_{DL_j}$. Set $Z_j = \text{Com}_{DL_j}(0, \tilde{r}'_j)$. Run $S_{HV}$ on input $x$ to get the view of $\overline{V}$ for session $j$, say $(\text{CMT}_j, \text{CH}_j, \text{RSP}_j)$. Set $Msg' = (\text{CMT}_j, Z_j)$ and $\text{next} = (Msg', j, \gamma)$.

**Step 2d:** If $Msg = V_2$ message of session $j$, i.e. $Msg = \beta_j$, then find $\text{CH}_j$, $\text{RSP}_j$ and $\tilde{r}'_j$ in $M'_{n,i} \cup F_{n,i+1}$ and set $\alpha_j = \text{CH}_j \oplus \beta_j$. Set $\tilde{r}_j = \text{Open}_{DL_j}(0, \alpha_j, \tilde{r}'_j, b_j)$. Set $Msg' = (\alpha_j, \tilde{r}_j, \text{RSP}_j)$ and $\text{next} = (Msg', j, \gamma)$.

**Step 2e:** If $Msg = \epsilon$, then set $Msg' = (p_\ell, q_\ell, g_\ell, g_\ell^{a_\ell})$ and $\text{next} = (Msg', \ell, \gamma)$.

**Step 2f:** If $Msg = (\text{END}, \text{output})$, $F_{n,i+1}$ outputs (END, output).

**Step 3:** Run $V^*$ on next and go to Step 2.

With the above description of $F_{n,i+1}$ complete, we now define $\{C_{n,i+1}, M_{n,i+1}, B_{n,i+1}, \mathsf{aux}_n\} = (C'_{n,i} \cup F_{n,i+1}, M'_{n,i} \cup F_{n,i+1}, B'_{n,i}, \mathsf{aux}_n)$.

Now that we have defined $\{C_{n,i+1}, M_{n,i+1}, B_{n,i+1}, \mathsf{aux}_n\}$, we define our admissible adversary $\{C'_{n,i+1}, M'_{n,i+1}, B'_{n,i+1}, \mathsf{aux}_n\}$ as follows: By $m$-KEA, there must exist a circuit $E_{n,i+1}$ which takes a subset of internal wires of $M_{n,i+1}$ as input and outputs $(j, b_j)$ such that if $X_j = B_j^{a_j}$ then $B_j = g^{b_j}$ with all but negligible probability. Here again, without loss of generality, for ease of notation, we assume that $E_{n,i+1}$ also outputs $j$ along with $b_j$. This can be done by just using output wires of $M_{n,i+1}$. Then, we define $\{C'_{n,i+1}, M'_{n,i+1}, B'_{n,i+1}, \mathsf{aux}_n\} = (C_{n,i+1} \cup E_{n,i+1}, M_{n,i+1}, B_{n,i+1} \cup E_{i+1}, \mathsf{aux}_n)$.

Now that we have defined this sequence of admissible adversaries, we will describe our simulator $\mathcal{S}_{V^*}$ in terms of these machines.

**Circuit:** $\mathcal{S}_{V^*}$.
**Input:** $(x, y)$, where $x \in L$ and $y \in \mathsf{aux}_n$ is the auxiliary input of length $m$.
**Output:** View of $V^*$.

**Step 1:** If $V^*$ starts $m$ sessions then $\mathcal{S}_{V^*}$ generates $(p_i, q_i, g_i) \xleftarrow{\$} L_{PQG}$ for all $i \in \{1, 2, \ldots, m\}$.

**Step 2:** $\mathcal{S}_{V^*}$ generates $a_1, a_2, \ldots, a_m$ uniformly at random such that $a_i \in \mathbb{Z}_{q_i}^*$ and computes $A_i = g_i^{a_i}$ for all $i$.

**Step 3:** $\mathcal{S}_{V^*}$ executes the admissible adversary circuit $(C_{n,m+1}, M_{n,m+1}, B_{n,m+1})$ with the inputs $(x, y, (p_1, q_1, g_1, g_1^{a_1}), \ldots, \ldots, (p_m, q_m, g_m, g_m^{a_m}))$ and $(R_1, R_2)$, where $x \in L$, $y$ is the auxiliary input of $V^*$ of length $m$ and $R_2$ is the random tape of $V^*$ and $R_1$ are the random coins for $C_{n,m+1}$.

**Step 4a:** If $(C_{n,m+1}, M_{n,m+1}, B_{n,m+1})$ outputs SimAbort then $\mathcal{S}_{V^*}$ also outputs SimAbort.

**Step 4b:** If $(C_{n,m+1}, M_{n,m+1}, B_{n,m+1})$ runs to completion with output (END, output), then $\mathcal{S}_{V^*}$ outputs output.

**Theorem D.11.** *If there are $m$ concurrent sessions of $\Pi$ and if our family of admissible adversaries $\mathcal{A}$ contains all polynomial size adversaries and allows polysize malicious extensions, then under $m$-KEA, honest verifier zero-knowledge property of $\langle \overline{P}, \overline{V} \rangle$ and the zero-knowledge property of $\Pi_{ZK}$, the following distribution ensembles are computationally indistinguishable:*

$$\{S_{V^*}(x,y)\}_{m, x \in L, y \in \{0,1\}^m} \text{ and } \{\langle P(x,w), V^*(x,y)\rangle\}_{m, x \in L, w \in W_L(x), y \in \{0,1\}^m}$$

**Proof:** We will prove indistinguishability by a sequence of hybrids. If there are $m$ sessions we will consider $3m + 1$ hybrids, $\mathcal{H}_0 \cup \{\mathcal{H}_{i,1}, \mathcal{H}_{i,2}, \mathcal{H}_{i,3}\}$, for all $i \in [m]$. We will now describe the hybrids in detail. We will assume that all the hybrids also have the witness $w$ for the fact $x \in L$.

- $\mathcal{H}_0$ is the honest hybrid. It runs Step 1 and 2 of $\mathcal{S}_{V^*}$ and then builds $F_{n,1}$ but does not stop on receiving the first response from $V^*$. Instead it uses the witness and interacts honestly in all the sessions. $\mathcal{H}_0 = \{F_{n,1}, F_{n,1}, \epsilon, \mathsf{aux}_n\}$. This hybrid is same as the honest prover interacting with $V^*$.

    In each of the following hybrids we build on the admissible adversary circuit $\{C'_{n,i}, M'_{n,i}, B'_{n,i}, \mathsf{aux}_n\}$. If its malicious part $M'_{n,i}$ outputs a tuple, we will call its session number $j$.

- $\mathcal{H}_{i,1}$ runs Step 1 and 2 of $\mathcal{S}_{V^*}$ and then builds $\{C'_{n,i}, M'_{n,i}, B'_{n,i}, \mathsf{aux}_n\}$ with the inputs $(x, y, (q_1, g_1, g_1^{a_1}), (q_2, g_2, g_2^{a_2}), \ldots, (q_m, g_m, g_m^{a_m}))$ and $(R_1, R_2)$, where $x \in L$, $y$ is the auxiliary input of $V^*$ of length $m$ and $R_2$ is the random tape of $V^*$ and $R_1$ are the random coins of $\{C'_{n,i}, M'_{n,i}, B'_{n,i}, \mathsf{aux}_n\}$. If $\{C'_{n,i}, M'_{n,i}, B'_{n,i}, \mathsf{aux}_n\}$ outputs SimAbort, then $\mathcal{H}_{i,1}$ does the same. Else find the last output from $V^*$ in $M_{n,i}$. It would be of the form $(j, B_j, X_j, \gamma)$. Now, start building polysize malicious extension $F_{n,i+1}$ to $\{C'_{n,i}, M'_{n,i}, B'_{n,i}, \mathsf{aux}_n\}$. $F_{n,i+1}$ continues as follows:

    - For the first $i$ sessions for which knowledge commitment was completed, whenever $\Pi_{ZK}$ completes do the following: If proof is not accepted, it adds (Abort, $\ell$) to Aborted. If proof is accepted, it finds the corresponding output $(\ell, b_\ell)$ of $B'_{n,i}$. If not found or

$B_\ell \neq g_\ell^{b_\ell}$, $F_{n,i+1}$ outputs SimAbort. If a valid $(\ell, b_\ell)$ is found and $\ell \neq j$, it uses the extracted values to simulate the sessions. For $\ell = j$, it acts honestly after $\Pi_{ZK}$.
- For rest of the sessions, it uses the witness to generate the messages honestly.

Note that though $\mathcal{H}_{i,1}$ has the extracted value for session $j$, it does not use it and acts honestly in that session. $\mathcal{H}_{i,1} = (C'_{n,i} \cup F_{n,i+1}, M'_{n,i} \cup F_{n,i+1}, B'_{n,i}, \mathsf{aux}_n)$.

- $\mathcal{H}_{i,2}$ is same as $\mathcal{H}_{i,1}$ with the following change. It chooses $\alpha_j \xleftarrow{\$} \mathbb{Z}_q$, but sets $Z_j = \mathsf{Com}_{DL_j}(0; \tilde{r}'_j)$. Later while opening it sets $\tilde{r}_j = \mathsf{Open}_{DL_j}(0, \alpha_j, \tilde{r}'_j, b_j)$ and opens the commitment to $\alpha_j$ and $\tilde{r}_j$. It generates all other messages of session $j$ honestly.

  The hybrids $\mathcal{H}_{i,1}$ and $\mathcal{H}_{i,2}$ are identical because the commitment scheme $\mathsf{Com}_{DL_j}$ is perfectly hiding and hence, $\mathsf{Com}_{DL_j}(0)$ and $\mathsf{Com}_{DL_j}(\alpha_j)$ are identically distributed.

- $\mathcal{H}_{i,3}$ does the following change in $\mathcal{H}_{i,2}$. While generating $P_2$ message of session $j$, it runs $S_{HV}$ to get $(\mathrm{CMT}_j, \mathrm{CH}_j, \mathrm{RSP}_j)$. It sends $(\mathrm{CMT}_j, \mathsf{Com}_{DL_j}(0; \tilde{r}'_j))$ as $P_2$ message. For $P_3$ message, it sets $\alpha_j = \beta_j \oplus \mathrm{CH}_j$, $\tilde{r}_j = \mathsf{Open}_{DL_j}(0, \alpha_j, \tilde{r}'_j, b_j)$ and sends $(\alpha_j, \tilde{r}_j, \mathrm{RSP}_j)$. Hence, in $\mathcal{H}_{i,3}$ all sessions $\ell$ such that $(\ell, b_\ell) \in$ output of $B^j_{n,i}$ are simulated and rest all sessions are honest. Note that $\mathcal{H}_{m,3}$ is same as the interaction between $\mathcal{S}_{V^*}$ and $V^*$.

  Below we will prove the indistinguishability of $\mathcal{H}_{i,2}$ and $\mathcal{H}_{i,3}$.
  First note that the only difference between the hybrids $\mathcal{H}_{i,3}$ and $\mathcal{H}_{i+1,1}$ is that in $\mathcal{H}_{i+1,1}$ we add $E_{n,i+1}$ to $B'_{n,i}$. But note that SimAbort can happen in $\mathcal{H}_{i+1,1}$ and not in $\mathcal{H}_{i,3}$ for the following two reasons:
  - When $V^*$ sends $(B, X)$ such that $(g_j^{a_j}, B, X) \notin \mathcal{DH}$ and manages to successfully complete the $\Pi_{ZK}$ by convincing that it sent a valid tuple. The extraction can fail almost always in this case. We prove in Lemma D.12 that probability of $V^*$ proving a wrong statement is negligible by reducing it to the statistical soundness of $\Pi_{ZK}$. Hence, the probability of SimAbort due to this event is also negligible.
  - When $V^*$ sends a valid $(B, X)$, yet $m$-KEA fails to give a successful extraction. But by the first property of $m$-KEA, the probability of this event is negligible in $n$.

Since the probability of each of the above events is negligible in $n$, $\mathcal{H}_{i,3}$ and $\mathcal{H}_{i+1,1}$ are statistically close. We now state and prove Lemma D.12 followed by indistinguishablity of $\mathcal{H}_{i,2}$ and $\mathcal{H}_{i,3}$.

**Lemma D.12.** Consider $\mathcal{H}_{i+1,1}$ such that $\mathcal{H}_{i,3}$ does not output SimAbort and $j$ as defined in $\mathcal{H}_{i+1,1}$. Then,

$$Pr[(g_j^{a_j}, B_j, X_j) \notin \mathcal{DH} \bigwedge \Pi_{ZK} \text{ is accepted for session } j] \text{ is negligible.}$$

**Proof:** We will prove this contradiction. If there is a non-negligible function $\gamma(n)$ such that $Pr[(g_j^{a_j}, B_j, X_j) \notin \mathcal{DH} \bigwedge \Pi_{ZK}$ is accepted for session $j] > \gamma(n)$ in $\mathcal{H}_{i+1,1}$, then we will construct an adversary, which will break the stand alone statistical soundness of $\Pi_{ZK}$ by $\gamma(n)$.

Consider a hybrid $\mathcal{H}'_{i+1,1}$ which is same as $\mathcal{H}_{i+1,1}$ with the following change: Internally, $\mathcal{H}_{i+1,1}$ extracts the secret value of $V^*$ for first $i+1$ sessions (where sessions are ordered according to knowledge commitment by $V^*$) and behaves honestly in all other sessions. For each of first $i+1$ knowledge commitments, it has an extractor circuit obtained from $m$-KEA. Since $\mathcal{H}_{i,3}$ does not output SimAbort, one of the following holds for each of the first $i$ extractions:

- The extraction is successful.
- $V^*$ fails to complete $\Pi_{ZK}$ successfully.

$\mathcal{H}'_{i+1,1}$ does not use any of the circuits output by $m$-KEA. Instead, it runs in super-polynomial time to extract the discrete log of $B_\ell$ w.r.t. $g_\ell$ corresponding to first $i$ knowledge commitments by $V^*$. This super-polynomial time extraction is always successful. Moreover, if $V^*$ fails to convince in $\Pi_{ZK}$ for some session, $\mathcal{H}'_{i+1,1}$ also aborts those sessions. For the sessions in which $\mathcal{H}_{i+1}$ acts honestly, $\mathcal{H}'_{i+1,1}$ also acts honestly. Hence, note that the view of $V^*$ is identical in $\mathcal{H}_{i+1,1}$ and $\mathcal{H}'_{i+1,1}$ till $V^*$ completes $\Pi_{ZK}$ for session $j$.

Let $V_{ZK}$ be an honest verifier for $\Pi_{ZK}$, which generates $(p,q,g) \xleftarrow{\$} L_{PQG}$ and $a \xleftarrow{\$} \mathbb{Z}_q^*$. Now consider a hybrid $\mathcal{H}''_{i+1,1}$ which is same as $\mathcal{H}'_{i+1,1}$ except for the change that it takes $(g, g^a)$ for the $j^{th}$ session from $V_{ZK}$. Since, $\mathcal{S}_{V^*}$ was generating these pairs at random and $V_{ZK}$ is honest, the view of $V^*$ in $\mathcal{H}''_{i+1,1}$ and $\mathcal{H}'_{i+1,1}$ is indistinguishable. Now, when $V^*$ sends $(j, B_j, X_j)$ for session $j$, $\mathcal{H}''_{i+1,1}$ forwards it to $V_{ZK}$. Whenever $V^*$ sends some message for $\Pi_{ZK}$ of session $j$, $\mathcal{H}''_{i+1,1}$ forwards it to $V_{ZK}$ and forwards the response of $V_{ZK}$ to $V^*$. Whenever $V^*$ convinces $\mathcal{H}''_{i+1,1}$, it also succeeds in convincing $V_{ZK}$. Since, $Pr[(g_j^{a_j}, B_j, X_j) \notin \mathcal{DH} \bigwedge \Pi_{ZK}$ is accepted for session $j] > \gamma(n)$ for $\mathcal{H}''_{i+1,1}$, it breaks the statistical soundness of $\Pi_{ZK}$ by probability $\gamma(n)$, which is a contradiction.

In order to show indistinguishability between $\mathcal{H}_0$ and $\mathcal{H}_{m,3}$, we are just left with showing indistinguishability between $\mathcal{H}_{i,2}$ and $\mathcal{H}_{i,3}$. We will show this by contradiction. Let us assume there is a distinguisher $\mathcal{D}$ which distinguishes between $\mathcal{H}_{i,2}$ and $\mathcal{H}_{i,3}$ for some $i$ and auxiliary input $y$. Then we will show a distinguisher $\mathcal{D}'$ for

$$\{S_{HV}(x)\}_{x \in L} \text{ and } \{\langle \overline{P}(x,w), \overline{V}(x) \rangle\}_{x \in L, w \in W_L(x)}.$$

This would contradict the honest verifier zero-knowledge property of $\langle \overline{P}, \overline{V} \rangle$. $\mathcal{D}'$ is given a 3-round transcript $\mathcal{T}' = \{\text{CMT}, \text{CH}, \text{RSP}\}$ as input for the NP-statement $x \in L$ which is either for $\overline{P}$ or $S_{HV}$ for the canonical argument system. Using the witness $w$ for $x \in L$, $\mathcal{D}'$ generates an input $\mathcal{H}_I$ for $\mathcal{D}$ which is same as $\mathcal{H}_{i,3}$ except for the following change. Instead of running $S_{HV}$ for the session $j$ (defined above), it uses $\{\text{CMT}, \text{CH}, \text{RSP}\}$ for that session.

If the input to $\mathcal{D}'$ is from $\{\langle \overline{P}(x,w), \overline{V}(x) \rangle\}_{x \in L, w \in W_L(x)}$, the input to $\mathcal{D}$ is identical to $\mathcal{H}_{i,2}$. This is because since honest verifier's CH is distributed uniformly in $\{0,1\}^n$, $\alpha_j$ in $\mathcal{H}_I$ will be distributed uniformly. On the other hand, if the input of $\mathcal{D}'$ is from $\{S_{HV(x)}\}_{x \in L}$, the input to $\mathcal{D}$ is identical to $\mathcal{H}_{i,3}$. So, if $\mathcal{D}$ says that $\mathcal{H}_I$ is distributed identically to $\mathcal{H}_{i,2}$, then $\mathcal{D}'$ says that $\mathcal{T}'$ is generated by $\overline{P}$. Else, it is generated by $S_{HV}$.

The success probability of $\mathcal{D}'$ in distinguishing between transcripts of $\overline{P}$ and $S_{HV}$ is same as the success probability of $\mathcal{D}$ in distinguishing $\mathcal{H}_{i,2}$ and $\mathcal{H}_{i,3}$. Hence, $\mathcal{D}'$ distinguishes between the following two ensembles with non-negligible probability

$$\{S_{HV}(x)\}_{x \in L} \text{ and } \{\langle \overline{P}(x,w), \overline{V}(x) \rangle\}_{x \in L, w \in W_L(x)}$$

in contradiction to property **B3** of the canonical argument system.